\documentclass[aps,pra,twocolumn,floatfix,superscriptaddress]{revtex4}
\usepackage{graphicx}
\usepackage{dcolumn}
\usepackage{float}
\usepackage{bm}
\usepackage{amssymb,bezier,epsfig,amsmath}
\usepackage{epsfig}
\usepackage{wrapfig}

\usepackage{txfonts}
\usepackage
{xspace}

\newcommand{\ket}[1]{\ensuremath{|#1\rangle}\xspace}
\newcommand{\bra}[1]{\ensuremath{\langle #1|}\xspace}
\input rotate
\input colordvi

\begin{document}

\title{Reaching the cold regime: S($^1$D) + H$_2$ and the Role of Long-Range Interactions in Open Shell Reactive Collisions}
\author{Manuel Lara} \email{mlaragar@gmail.com}
\affiliation{Institut de Physique de Rennes, UMR CNRS 6251,
Universit\'e de Rennes I, F-35042 Rennes, France}
\author{F. Dayou}  
\affiliation{{LERMA, UMR 8112 du CNRS, 
Observatoire de Paris-Meudon, Universit\'e Pierre et Marie Curie, 92195 Meudon Cedex, France}}
\author{J.M Launay}
\affiliation{Institut de Physique de Rennes, UMR CNRS 6251,
Universit\'e de Rennes I, F-35042 Rennes, France} 

\date{\today}

\begin{abstract}
Reactive cross-sections for the collision of open shell S($^1$D) atoms with {\em ortho}- and {\em para}-
hydrogen, in the kinetic energy range 1$-$120 K, 
have been calculated using the hyperspherical quantum reactive scattering method developed by Launay {\em et al.} [\textit{Chem. Phys. Lett.} {\bf 169}, 473 (1990)]. 
Short-range interactions, described using the {\em ab initio} potential energy surface by
Ho {\em et al.},
 were complemented with an accurate description of the long-range interactions, where the main electrostatic ($\sim R^{-5}$) and dispersion ($\sim R^{-6}$) contributions were considered. This allows the comparison with recent experimental measurements of rate constants and excitation functions for the title reaction at low temperatures [Berteloite {\em et al.}, accepted in \textit{Phys. Rev. Lett.}, 2010].
The agreement is fairly good. The behavior in the considered energy range can be understood on the average in terms of a classical Langevin (capture) model,
 where the centrifugal barriers determine the amount of reactive flux which reaches the barrierless transition state. Additionally,  the structure of the van der Waals well provides temporal trapping at short distances thus allowing the system to find its way to the reaction at some classically-forbidden energies. Interestingly, the cross-section for {\em para}-hydrogen shows clearly oscillating features associated to the opening of new partial waves and to shape resonances which may be amenable to experimental detection.  
\end{abstract}

\maketitle

\section{Introduction}
Until very recently, most experimental and theoretical studies in reaction dynamics had focused on thermal (or higher) collision energies. 
With the field of cold molecules emerging as an active interdisciplinary area of research, 
a great interest in understanding cold and ultracold chemical reactions has risen~\cite{coldcol,Fara}. Electric and magnetic decelerators, helium buffer gas cooling, and magneto- and photo-association (among other techniques) are opening the access to cold and ultracold samples of molecules~\cite{Bell}.
Given the success in controlling interactions achieved during the last two decades in the field of ultracold atomic physics~\cite{Hutsonrev}, the question is whether a similar level of control is possible for more general processes involving molecules. Very few partial waves contribute at low temperatures. However,
reactive processes have been  shown to occur rapidly being dominated 
by tunneling and Feshbach resonances, in a regime where the smallest barrier to reaction surely exceeds the collision energy~\cite{UltraD,elramsa,bodobodo,Weckeur,WeckphyB}. 
Remarkably, at low kinetic energies the presence of external fields modifies the dynamics in an essential way, appearing as ``knobs'' to steer the system to the desired outcome~\cite{Lihfelec}. In addition, the theoretical advances in the coherent control of bimolecular processes
 achieved in the nineties\cite{Shapi,Zeman} found in the thermal averaging of the colliding partners a fundamental obstacle.
Consequently, the forthcoming access to coherent samples of molecules at very low temperatures has renewed the quest for the control of chemical reactions\cite{FelipeH}.

 Given the success in the production of (ultra-)cold alkali metal dimers,
attention has mainly focused in such particular systems~\cite{Staanum,Zahzam,Wynar,Mukaiyama,Syassen,hudsonexp}.
In a landmark experiment, reaction rates for collisions involving  fermionic $^{40}$K$^{87}$Rb, at temperatures less than
1 $\mu$K, were recently measured~\cite{Ospel}. In contrast, advances in the analysis of other type of reactions are still hampered by  the lack of versatile methods to produce molecules at low temperatures and the very low densities achieved. Luckily, emerging experimental techniques like the Crossed Stark decelerators~\cite{FaraBas} promise to provide detailed information on cold bimolecular collisions.
More related with our work, CRESU (Reaction Kinetics in Uniform Supersonic Flow) experiments implemented in Rennes~\cite{Canosa,nuestroFara}, and crossed molecular beam experiments performed in Bordeaux~\cite{Costes,Costes2},
are exploring barrierless reactions between neutral species in conditions aproaching the cold regime~\cite{personal}. In particular, rate constants and cross-sections have been recently obtained for the title reaction at
temperatures and kinetic energies as low as $\sim 5$K (1K$\sim 8.3\times10^{-3}$KJ~mol$^{-1}$)~\cite{Science}. We will analyze their results below.

Given the availability of experimental results, the title reaction provides us with a good chance to check and extend
 our predictive power from the explored thermal regime, where {\em ab initio} reaction dynamics is routinely applied, to the low energy one 
where such methodology should fail due to the lack of accuracy. 
In the limit, the deep ultracold regime, governed by Wigner laws, has been found amenable to 
parametrization in  terms of a very few inputs (the scattering length, for example) 
which the experiments may provide soon~\cite{Ospel}. Paradoxically, the extreme sensitivity
of such parameters to small details of the PES, and the action of the surface as a whole, 
makes extremely difficult to predict them or, conversely, 
to deconvolute from their measurement the underlying interactions. 
 Collisions in the range of $\sim 1$ K thus lie in the limits of what can be done using our conventional theoretical tools. Besides, 
such collisions are more than an obliged milestone 
in our route from the thermal to the ultracold regime, 
having interest on their own. They report the combined game of short-range (SR) and long-range (LR) interactions.
  While the ultracold scenario  
privileges  LR interactions, leading to universality in
 extreme cases\cite{Fara}, the thermal regime  and its higher kinetic 
energies makes (SR) chemical forces prevail. As we will show, 
both SR and LR interactions play the game at the concerned kinetic energies. 
Nevertheless, their roles may be independent enough to allow to associate 
trends in the behavior to  different regions of the PES, thus offering some insight
in the underlying dynamics.

 The S($^1$D) + H$_2(X\ ^1\Sigma_g^+)$ $\rightarrow$ SH + H  thermal reaction and its isotopic variants were the subject of detailed analysis in the past~\cite{Skodje01a,Skodje02,Banares02,Banaresulti,HonvaLau,Rack1,Tomas,klos,Maiti,Liu98a,Liu98b,Liu00}.
 Belonging to the class of {\em insertion reactions}~\cite{aoizfe}, the system is characterized by the presence of a deep well (96 kcal mol$^{-1}$) on its singlet ground potential energy surface (the 1$^1A'$ electronic state of H$_2$S) and by a small exoergicity (6.90 kcal mol$^{-1}$).
Lee and Liu carried out
crossed molecular beam experiments which were able to conclude the insertion mechanism~\cite{Liu98a,Liu98b,Liu00}. 
Their measurements motivated a number of theoretical studies.
Zyubin {\em et al.}~\cite{Skodje01a}, working at a
multi-reference configuration interaction (MRCI) level with multi-configuration
self-consistent field (MCSCF) reference wave functions,
computed the five (1A', 2A', 3A', 1A'' and 2A'') potential energy surfaces (PES)
which correlate with the S($^1$D) + H$_2$ asymptote.  
 The collinear barrier 
of the first  1$^1$A'' excited PES was
found to be 10 kcal mol$^{-1}$ high. The other three surfaces being repulsive,  the reaction at moderate energies may be assumed to  occur on
the 1A' PES. An improved ground state, based on the same {\em ab initio} data set was subsequently produced by Ho 
 {\em et al.}\cite{Skodje02}. This PES is the one we will use and it will be labeled as S1. Working on it, the quantum calculations of Honvault and Launay~\cite{HonvaLau},
 precursors to the ones we will show below, constituted landmark single-PES
calculations on atom-diatom
 insertion reactions at thermal energies. 
 Relevant to the discussion of our results will be the statistical approaches 
to the system~\cite{Changg,Rack1,Aoiztomas,Tomas,Ying}. Let us recall the statistical model of Rackham and Manolopoulos~\cite{Rack0} and the quasiclassical statistical model~\cite{Aoiz3}. In Refs.~\cite{Rack1} and ~\cite{Aoiz3} the latter were shown to account very well for the QM results by Honvault and Launay. More recently, non-adiabatic effects in the product channel were introduced in 
the quantum model in order to calculate reaction cross-sections resolved in rotational and 
fine-structure product states~\cite{klos}. 

  Much lower in energy in the reagent region there exist electronic surfaces which correlate to the S($^3 P$), cross the $^1A'$ surface in the  H$_2$S well and lead to the same asymptote in the product valley. The role of the intersystem crossing in the S($^3$P,$^1$D) + H$_2$ collisions was analyzed  by Maiti {\em et al.}\cite{Maiti} using trajectory surface-hopping methodology. Their calculations conclude that the electronic
 quenching process, S($^1$D) + H$_2$ $\rightarrow$ 
S($^3$P$_{0,1,2}$) + H$_2$, plays a major role in the removal of S($^1$D) at energies $\sim 250$K. 
 The significance of non-adiabatic effects in the title system was also revealed in a recent quantum study
 performed on the isotopic variant  S($^1$D) + HD~\cite{Tian}. We thus expect a significant contribution of the quenching process in the current regime. 

 In this work we will show the results of new quantum mechanical calculations for the title collision at very low temperatures. They were obtained starting from an accurate 
description of the LR interactions, essential in order to describe  collisions at low energy\cite{Weck2006}. In a regime where the kinetic energy is so small, LR interactions determine the amount of incoming flux which reaches the SR, where rearrangement may occur\cite{Ospel}. In our case the system is  characterized by the presence of a significant quadrupole-quadrupole (QQ) contribution, which should lead to important reorientation effects. Therefore, the dynamical methodology has been revisited in order to include the most of the anisotropy that an adiabatic electronic treatment may enable.  The theoretical results will be compared with the experimental results from the work by Costes and co-workers and Sims and co-workers~\cite{Science}.

 The paper is structured as follows. In the next section, we will briefly describe the new LR calculations, while further details will be given in the Appendix. In section III, the hyperspherical approach to the dynamics\cite{Launayfirst} will be recalled and small methodological improvements will be explained. The results from the dynamical calculations will be shown in section IV and they will be discussed and compared with the experimental data in Section V. Finally, a summary of the work and the conclusions will be given in Section VI.

\section{Long-range interactions}
\label{longrange}

\subsection{Formalism}

The LR interactions between the sulfur atom S$(^1{\rm D})$ and hydrogen molecule H$_2(X\ ^1\Sigma_g^+)$ have been described following perturbation theory up to second order, using a multipolar expansion of the electrostatic interaction operator ${\hat H}_{el}$. To account for the 5-fold degeneracy of the open-shell atom S$(^1D)$, we defined a set of asymptotically degenerate diabatic states, taken as a product of the atomic $\ket{L\lambda}^{(A)}$ and diatomic $\ket{0}^{(B)}$ unperturbed electronic states.  The quantum number $\lambda=0,\pm 1,\pm 2$ is the projection of the atomic orbital angular momentum ${\bf L}$ along the Body-Fixed (BF) $z$-axis, chosen along the intermolecular vector ${\bf R}$, and the projection $\lambda=0$ of the diatomic orbital angular momentum relates to the diatom axis ${\bf r}$, chosen to be in the BF $xz$-plane. Following previous works on open-shell systems~\cite{bussery:08,groenenboom:06,spelsberg:99,zeimen:03}, the matrix elements of the first-order and second-order perturbation operators in the $\ket{L\lambda}^{(A)}\ket{0}^{(B)}$ basis set lead to LR potentials that depend on the internal coordinates $(R,\theta)$, with $R$ being the intermolecular separation and $\theta$ the angle between ${\bf R}$ and ${\bf r}$. The dependence of the LR potentials on $r$, the internuclear distance of H$_2$, has been neglected in the present case. We have fixed $r$ at its vibrationally averaged value in the ground state, $\left<r\right>_{v=0}$.

The anisotropy of the LR S$(^1{\rm D})$-H$_2$ interaction can induce transitions between the fragment states $\ket{L\lambda}^{(A)}\ket{0}^{(B)}$ associated with different values of the projection quantum number $\lambda$, whereas $L$ is assumed to be conserved throughout the collision for large $R$. We can therefore drop the quantum number $L$ and label by $\ket{\lambda}$ the electronic diabatic basis of the S$(^1{\rm D})$-H$_2$ interacting system. For linear geometries ($\theta=0$), there is one-to-one correspondence between the states $\ket{\lambda}$ and the adiabatic states of $\Sigma$, $\Pi$ and $\Delta$ symmetry of S$(^1{\rm D})$-H$_2$. Those adiabatic states are labeled according to the value of the quantum number $\Lambda=0,\pm 1,\pm 2$, the projection of the total electronic orbital angular momentum ${\bf\cal L}$ along the BF $z$-axis, with ${\bf\cal L}$ being defined with respect to the triatomic center-of-mass BF frame. For non-linear geometries, the diabatic states $\ket{\lambda}$ are coupled through LR interactions, and one has to deal with a $5\times 5$ LR potential matrix $V_{\lambda,\lambda'}(R,\theta)$. Since the issue of non-adiabatic couplings is not addressed in present work, we chose to keep the (complex) signed-$\lambda$ diabatic basis $\ket{\lambda}$ to express the LR potential matrix elements. The diagonalization of the  $5\times 5$ LR potential matrix yields the adiabatic potentials associated with the three $A'$ and two $A''$ adiabatic states correlating with S$(^1{\rm D})$-H$_2$. 

In the present study, we have considered the electrostatic interaction between the permanent quadrupole moments of S$(^1{\rm D})$ and H$_2$, leading to electrostatic energies varying as $R^{-5}$, and the dispersion interaction between the dipole-induced moments of the two species, giving rise to dispersion energies varying as $R^{-6}$. The contribution of induction effects to the LR potentials has been neglected, since the dominant induction contribution involves interactions between permanent quadrupole moments and dipole-induced moments, which are proportional to $R^{-8}$. According to Eqs.~\ref{elec} and~\ref{disp} of the Appendix, the LR potential matrix elements write:

\begin{equation}
\label{LR}
V_{\lambda,\lambda'}(R,\theta) =  \frac{1}{R^5}C_5^{\lambda\lambda'}\;C_{2,\lambda'-\lambda}(\theta,0)
-\frac{1}{R^6}\sum_{k=0,2}C_{6,k}^{\lambda\lambda'}\;C_{k,\lambda'-\lambda}(\theta,0)
\end{equation}

\noindent  where the angular functions $C_{l,m}(\theta,\phi)$ are normalized spherical harmonics\footnote{The normalized spherical harmonics  $C_{l,m}(\theta,0)$ relate to the associated Legendre polynomials $P_l^m(\cos\theta)$ as follows: $$C_{l,m}(\theta,0)=(-1)^m\left[\frac{(l-m)!}{(l+m)!}\right]^{1/2}P_l^m(\cos\theta)$$ \noindent for $m\ge 0$, and $C_{l,m}(\theta,0)=(-1)^mC_{l,|m|}(\theta,0)$ for $m<0$.}, and the interaction coefficients $C_5^{\lambda\lambda'}$ and $C_{6,k}^{\lambda\lambda'}$ stands for the electrostatic (quadrupole-quadrupole) and dispersion (dipole-induced dipole-induced) contributions, respectively.

\subsection{Quadrupole moments and dipole polarizabilities}

The evaluation of the LR electrostatic $C_5^{\lambda\lambda'}$ and dispersion $C_{6,k}^{\lambda\lambda'}$ coefficients requires to determine the permanent quadrupole moments and dynamic dipole polarizabilities of S$(^1{\rm D})$ and H$_2$. For H$_2$, we have selected accurate values from the literature, corresponding with a H$_2$ geometry at its vibrationally averaged value, $\left<r\right>_{v=0}=1.449$ $a_0$. We used the permanent quadrupole value $\bra{0}{\hat Q}_{20}\ket{0}=+0.481$ a.u. of Ref.~\cite{spelsberg:93}, obtained from full Configuration Interaction (CI) calculations with a large Gaussian-type orbitals basis set, including bond-centered polarization functions. For the dynamic dipole polarizabilities $^{00}\alpha_{zz}(i\omega)$ and $^{00}\alpha_{xx}(i\omega)$, we chose the values tabulated in Ref.~\cite{bishop:92}, obtained from a sum-over-states formalism with explicitely electron-correlated wavefunctions. The corresponding static polarizabilities, $^{00}\alpha_{zz}(0)=6.721$ a.u. and $^{00}\alpha_{xx}(0)=4.739$ a.u., are in good agreement with other literature values~\cite{spelsberg:93}.

For S$(^1D)$, we performed {\em ab initio} calculations by means of the Dalton quantum chemistry code~\cite{dalton}. The permanent quadrupole moments have been calculated as the expectation values of cartesian quadrupole moment operators, $\hat{\Theta}_{uu}$ with $u=\{x,y,z\}$, using {\it ab initio} MCSCF electronic wavefunctions. The MCSCF wavefunctions were generated by distributing six electrons among 13 orbitals ($3s$, $3p$, $3d$, $4s$ and $4p$), the inner-shell orbitals being kept frozen in their ROHF form, and the Sadlej-$p$VTZ basis set was used. The resulting value of the quadrupole moment $\bra{L0}\hat{Q}_{20}\ket{L0}=+2.075$ a.u. was employed to get the whole set of nonzero matrix elements from the Wigner-Eckart theorem. Notice that the latter value is identical to the one already reported in the literature~\cite{medved:00,andersson:92} for the same level of calculation. The dynamic dipole polarizabilities of S$(^1D)$ have been determined by means of MCSCF linear response calculations, using the MCSCF wavefunctions previously defined. For each substate of definite symmetry, we obtained a set of Cauchy moments, each of them corresponding with a given cartesian component of the electric dipole operator $\hat{\mu}_{u=x,y,z}$. We then employed analytical continuation techniques following the $\left[n,n-1\right]_{\alpha}$ and  $\left[n,n-1\right]_{\beta}$ Pad\'e approximants procedure of Ref.~\cite{langhoff:70}  to get lower and upper bounds to the dynamic polarizabilities $^{vv}\alpha_{uu}(i\omega)$, where $v$ stands for an atomic substate of definite symmetry. For $n=9$, the dispersion coefficients associated to lower and upper bounds of the polarizabilities are converged within less than 0.02$\%$. The values of the corresponding static polarizabilities are identical to those already published~\cite{medved:00}. From the diagonal elements of the dynamic polarizabilities $^{vv}\alpha_{uu}(i\omega)$, we derived the whole set of coupled spherical dynamic polarizabilities $^{\lambda\lambda'}\alpha_{(11)kq}$ following the procedure described in Ref.~\cite{bussery:08}.

The resulting LR interaction coefficients for the electrostatic $C_5^{\lambda\lambda'}$ and dispersion $C_{6,k}^{\lambda\lambda'}$ contributions are tabulated in the electronic supplementary information (ESI). To the best of our knowledge, no LR coefficients had been determined so far for the S$(^1D)$-H$_2$ system, and thus it is not possible to compare with the results of previous works. By summing the tabulated values over the diagonal matrix elements ($\lambda=\lambda'$) we can retrieve state-averaged interaction coefficients. The state-averaged electrostatic contribution vanishes, and we get for the isotropic and anisotropic dispersion contributions $\bar{C}_{6,0}=40.338$ a.u. and $\bar{C}_{6,2}=4.146$ a.u., respectively.

\section{Dynamical Methodology }

\subsection{The hypersperical approach}

The quantum methodology used to carry out the dynamical calculations
was described in previous works on 
alkalis at ultracold energies~\cite{Sol02,Quem04}, and elsewhere in the context of
thermal reactive scattering\cite{hon04}. In fact, most of the convergence parameters used in this work are the same which
were used in the study of the title collision at thermal 
energies\cite{Banaresulti}.
 Let us simply recall that in the hyperspherical quantum reactive scattering method developed by
J. M. Launay\cite{Launayfirst} the configuration space is divided into
inner and outer regions. The positions of the nuclei in the inner region 
are described in terms of hyperspherical
democratic coordinates. The logarithmic derivative of the wavefunction is propagated outwards on a single adiabatic PES. At a large enough value of the hyper-radius the former is matched to a set of suitable functions, called asymptotic functions, to yield the scattering S-matrix. We chose an intermolecular separation of $\approx 10$ a.u. for the matching.

  The asymptotic functions provide the collisional boundary conditions. When working at thermal energies they 
are familiar regular and irregular radial Bessel functions
which account for the presence of the centrifugal 
potential at large intermolecular separation. Recently, they were propertly modified to include the effect of an 
isotropic dispersion interaction ($\sim R^{-6}$), thus enabling the study of ultracold collisions of 
alkalis\cite{Sol02,Quem04}. In the current work further changes are required. Asymptotic 
functions must account also for the presence of the 
anisotropic electrostatic and dispersion LR interaction, introducing reorientation effects which may take place while the reagents approach. 
 More details about their definition are given in the following section.

\subsection{Introducing the anisotropy in the outer region:  the asymptotic functions}\label{efe}
Using the set of Jacobi coordinates $(R,r,\theta)$ corresponding to the
S+H$_2$ arrangement, the total Hamiltonian of the system can be expressed as 
\begin{eqnarray}
{\hat H} =- \frac{\hbar^2}{2 \mu} \frac{1}{R} \frac{\partial^2}{\partial
R^2} R  +\frac{1}{2 \mu
R^2} {\bf l}^2- \frac{\hbar^2}{2 m} \frac{1}{r} \frac{\partial^2}{\partial
r^2} r  +\frac{1}{2 m
r^2} {\bf j}^2 + {\hat H}_{el}  
\label{hamil}\end{eqnarray}
\noindent where ${\hat H}_{el}$ is the electronic Hamiltonian, ${\bf l}$ is the orbital angular momentum
of the atom with respect to the center of mass of the diatom, and ${\bf j}$ is the
rotational angular momentum of the latter. The total
angular momentum of the nuclei (conserved in an adiabatic approach) is given by $\bf{J}=\bf{j}+\bf{l}$. In the current study, we chose an adiabatic treatment of the dynamics, assuming that the collision occurs only on the ground adiabatic PES. Hence, among the five adiabatic PESs correlating with S($^1$D)+H$_2$ (three $A'$ and two $A''$ singlet states), we consider only the lowest one, associated with the ground $1\ ^1A'$ electronic state. Hereafter, the lowest eigenvalue of ${\hat H}_{el}$ will be labeled by $V^0(R,r,\theta)$. The diagonalization of the LR potential matrix of Eq.~\ref{LR} provides a value for such energy, accurate at large intermolecular separations, with $r$ fixed to its vibrationally averaged value $\left<r\right>_{v=0}$. 

A convenient basis in order to expand the nuclear wavefunction in the LR region is the one characterized by  quantum numbers $(J,M,v,j,l)$, with ($v,j$) the rovibrational quantum numbers of the diatom, $l$ the  relative orbital angular momentum and ($J,M$) the total angular momentum and its projection on the Space-Fixed $Z$ axis. We will call this basis B1, and represent it as $\phi^{J M}_{v j l }$. 
Such a basis, well adapted to handle Coriolis couplings, is used in the hyperspherical approach to expand the asymptotic wavefunctions, which are matched with the SR information obtained in hyperspherical coordinates.

If the system approaches collision with quantum numbers $(J,M,v_0,j_0,l_0)$, we will asumme
 that (in addition to $J$ and $M$) the rovibrational quantum numbers, ($v_0,j_0$), remain well conserved in the LR region. This is well justified in the present case given the large energy gap between different rovibrational states relative to the considered collision energy range. 
Within this approximation the total electro-nuclear wavefuncion, $\Psi^{J M}_{v_0 j_0 l_0 }$, can be expanded in the LR region as  
\begin{eqnarray}\label{eq:desar2} 
\Psi^{J M}_{v_0 j_0 l_0 }=\sum_{l} \frac{F^{l_0}_l(R)}{R} \phi^{J M}_{v_0 j_0  l}  | 1\  ^1A' \rangle 
\label{expa}
\end{eqnarray}
\noindent where $| 1\  ^1A' \rangle $ is the ground adiabatic electronic state and all the "conserved" quantum numbers have been suppressed in the notation of the radial coefficients, $F^{l_0}_l(R)$. Introducing the expansion
 (\ref{expa}) into the time-independent Schr{\"o}dinger equation associated with a total energy $E$, $H \Psi=E \Psi$, and using the Hamiltonian in Eq.~\ref{hamil}, it is straightforward to obtain the following system of coupled radial equations
 \begin{eqnarray}\label{eq:acoplamien}
 \left[ - \frac{\hbar^2}{2 \mu}
 \frac{\partial^2}{\partial
R^2}  + \frac{l(l+1) \hbar^2 }{2 \mu R^2}  -E_c \right] F^{l_0}_{l}(R)=  
- \sum_{ l' } \langle  V \rangle_{ l, l'} (R) F^{l_0}_{ l'}(R) 
\end{eqnarray}
\noindent where $E_c$, the collision energy, is given by  $E-E_{v,j}$, with $E_{v,j}$ the internal energy of the diatom. $\langle V \rangle _{ l, l'} (R)$ designs the matrix elements in basis B1 of  $V^0(R,r, \theta)-V_{H_2}(r)$, with $V_{H_2}(r)$ the asymptotic H$_2$ diatomic potential. 
By inwards integration of Eq.~\ref{eq:acoplamien} we obtain the 
``regular'' ($F^{(1) l_0}_{l}$)
 and ``irregular'' ($F^{(2)l_0}_{l}$) asymptotic radial wavefunctions, corresponding to an incoming $(J,M,v_0,j_0,l_0)$ channel. They are defined as
the ones  which behave as
 \begin{eqnarray}\label{eq:desar2}
F^{(1)l_0}_{ l}  \stackrel{R \rightarrow \infty}
{\longrightarrow} \sin(kR-l \pi /2) \delta_{l l_0} / k^{1/2}   \\ \nonumber
F^{(2)l_0}_{l}  \stackrel{R \rightarrow \infty}
{\longrightarrow} -\cos(kR-l \pi /2) \delta_{l l_0} / k^{1/2} 
\end{eqnarray} 
\noindent where $k$ is the wavenumber of the considered channel.
Such radial functions account for the correct LR behaviour
 to match with the inner-dynamics in the presence of an anisotropic $V^0(R,r, \theta)$ potential. 
The  coupled-equation version of the method of
 De Vogelaere \cite{Lester} is used to solve for them.

Regarding the calculation of the potential matrix elements $\langle V \rangle _{ l, l'} (R)$, it is convenient to define an intermediate nuclear basis, B2, labelled by the projection $\Omega_j$ of  $\bf{j}$ and  $\bf{J}$ on the BF $z$-axis. The basis set B2 is given by
\begin{eqnarray}\label{eq:jotas}
\phi^{J M}_{v  j  \Omega_j }=\frac{\chi_{v,j}(r)}{r}
\sqrt{\frac{2J+1}{4\pi}}D_{M \Omega_j}^{J*}(\alpha,\beta,\gamma)Y_{j
  \Omega_j}(\theta,0), 
\end{eqnarray}
\noindent where $D_{M \Omega_j}^{J*}$ designs a Wigner rotation matrix element and $(\alpha,\beta,\gamma)$ are the Euler angles corresponding to the transformation between SF and BF frames. The change of basis between B1 and B2 simply involves 3j symbols. The matrix elements of the ground PES in basis B2 
are given by 

\begin{equation}\label{clarito}
\langle  V \rangle_{\Omega_j, \Omega_j'}(R)= \delta_{\Omega_j  \Omega_j'} 2 \pi  \int  \chi_{v,j}^2(r)
 Y_{j \Omega_j}^2(\theta,0) V^0(R,r,\theta) \sin\theta\; dr\; d\theta
\end{equation}

\noindent 
 The potential 
does not couple states with different $\Omega_j$, although reorientation effects in the BF system happen due to the Coriolis term.  Assuming in Eq.~\ref{clarito} that the variations with $R$ of the mean value of $V^0(R,r,\theta)$ in the rovibrational state $\chi_{v,j}(r)$ are well approximated by the variations of the potential at the averaged bond distance, $r=\left<r\right>_{v,j}$ (assumption which has been proven valid for the current PES), we reach the convenient expression
\begin{widetext}
\begin{eqnarray}\label{aproximo}
\langle  V \rangle_{\Omega_j, \Omega_j'}(R) 
\approx  \delta_{\Omega_j  \Omega_j'} 2 \pi  \int
 Y_{j \Omega_j}^2(\theta,0)~ V^0(R,\left<r\right>_{v,j},\theta)  \sin\theta\; d \theta.
\end{eqnarray}
\end{widetext}
\noindent Once the potential matrix elements are calculated in basis B2 we change to basis B1, thus obtaining the elements $\langle  V \rangle_{ l, l'} (R)$ involved in Eq.~\ref{eq:acoplamien},
\begin{eqnarray}\label{camio}
\langle  V \rangle_{l,l'}(R) & = & (-1)^{l+l'} \sqrt{2l+1}\sqrt{2l'+1} \\
&&\times  \sum_{\Omega_j} \left(\begin{array}{ccc} j & l & J  \\ \Omega_j & 0  & -\Omega_j \end{array}\right)
\left(\begin{array}{ccc} j & l' & J  \\ \Omega_j & 0  & -\Omega_j \end{array}\right)  \langle  V \rangle_{\Omega_j, \Omega_j}(R). \nonumber
\end{eqnarray}
\noindent Let us finally note  that $\langle  V \rangle_{l_0,l_0}(R)+ l_0(l_0+1) \hbar^2 /2 \mu R^2$ has the meaning of the effective potential felt by the colliding partners at a distance $R$ when approaching in the state $\phi^{J M}_{v_0  j_0 l_0}$. Such meaning will be used below. 

\subsection{The potential energy surface}

 In our adiabatic approach we consider that the collision takes place on the ground PES. The global fit for the $1^{1}A'$ PES performed by Ho {\em  et al.}\cite{Skodje02} (surface S1) was widely used in the past and its SR region was tested by comparison with experiments at higher energies \cite{Banares02,Banaresulti}. Nevertheless, the surface does not pay the necessary attention to the description of the LR region, that being the reason why we performed our own  calculations of the LR potentials (see Sec.~\ref{longrange}).
 To combine the SR interactions given by S1 with our description of the LR region (the lowest eigenvalue of the LR potential matrix in Eq.~\ref{LR}), we performed a smooth switching from one to the other. Thanks to the approximation in Eq.~\ref{aproximo}, only the LR potentials for fixed  $r= \left<r\right>_{v,j} \sim \left<r\right>_{v=0}$ are required, and only the switching for such internuclear distance is considered. The switching was performed around the central value $R_0=12.5$ a.u., within a range of intermolecular distances given by $\Delta R = 3$ a.u.  The angle-independent switching function $f(R)=\{\cos[\pi( R+ \Delta R/2 -R_0)]+1\}/2 $ was used. 
 In contrast, note that an intermolecular  separation of $\approx 10$ a.u. was considered for the matching of the inner dynamics with the external asymptotic functions. 

It is worth mentioning that a new global $1^{1}A'$ PES for the title system was recently published~\cite{Var}. A term to account for the ($\sim R^{-6}$) dipole-induced dipole-induced (DD) dispersion interaction was included in the global fit. Nevertheless, the  quadrupole-quadrupole (QQ) electrostatic interaction ($\sim R^{-5}$), which dominates the LR behavior, was not explicitly considered. As such term is relevant to describe the reactivity of {\em ortho-}hydrogen at low collision energies, we have preferred to employ the PES obtained from the switching procedure just described.


 \subsection{The collision with j=0 {\em  para}-hydrogen
\label{para}}
The collision of S($^1$D) with {\em para-}hydrogen in its ground rotational state deserves special attention. Within a non-adiabatic treatment the effective
 contribution of the anisotropic terms in $V_{\lambda,\lambda'}(R,\theta)$ (Eq.~\ref{LR}) is found to vanish. This is not the case when using an adiabatic approach instead. In order to show this we define a set of electro-nuclear 
basis functions in terms of the diabatic electronic basis defined in Section \ref{longrange},
\begin{eqnarray}\label{eq:jotas2}
\phi^{J_T M_T}_{v  j  \Omega_j L \lambda}=\frac{\chi_{v,j}(r)}{r}
\sqrt{\frac{2J_T+1}{4\pi}}D_{M_T \Omega_T}^{J_T*}(\alpha,\beta,\gamma)Y_{j
  \Omega_j}(\theta,0) | L \lambda \rangle
\end{eqnarray}
 where $J_T$  designs a total 
angular momentum, $\bf{J_T}=\bf{j}+\bf{l}+\bf{L}$, which in contrast with Eq.~\ref{eq:jotas} includes the atomic electronic angular momentum, 
 and  $\Omega_T$ ($= \Omega_j + \lambda $) and $M_T$ label its projection on the BF and SF $z-$axis, respectively.
 The use of this basis in combination with the whole interaction matrix given by Eq.~\ref{LR}  would efficiently account for the exchange of angular momentum among electrons and nuclei ($J_T$ being conserved) and reorientation  effects appearing while the reagents approach~\cite{Druk}.
Potential couplings between two different channels
in this non-adiabatic treatment would involve the integration in $\theta$  of the product of two spherical harmonics (originating from Eq.~\ref{eq:jotas2}) and  one matrix element 
$V_{\lambda,\lambda'}(R,\theta)$. The latter consists for spherical harmonics, $Y_{k,m}$,  with $k=0, 1, 2$ ($C_{k,m} \sim Y_{k,m}$ in Eq.~\ref{LR}).
 When considering the case of {\em para-}hydrogen with j=0, such integral is found to vanish for $k \ne 0$. The contribution of  QQ and dispersion anisotropic terms disappears, and thus the $ R^{-5}$ dependence.
Only the isotropic dispersion term, $\sim R^{-6}$, is found to play a role. 
This is not the case when the adiabatic approach is used. The diagonalization of the LR $V_{\lambda,\lambda'}(R,\theta)$ matrix yields a  ground PES with a  $R^{-5}$ dependence, which is not cancelled through the integral in Eq.~\ref{aproximo}.
In order to correct such a drawback of the adiabatic treatment we artificially matched S1  with a pure 
isotropic dispersion term,  $-C_6/ R^{6}$, when dealing with  j=0 {\em para-}hydrogen. 
In this way we introduced in the adiabatic treatment the effective LR 
interaction which the colliding partners should experience.  The $C_{6,k=0}^{0,0}$ coefficient, the lowest isotropic one, was chosen for that purpose.

\section{Results}
\begin{figure}
\setlength{\unitlength}{8mm}
\begin{center}
\epsfig{file=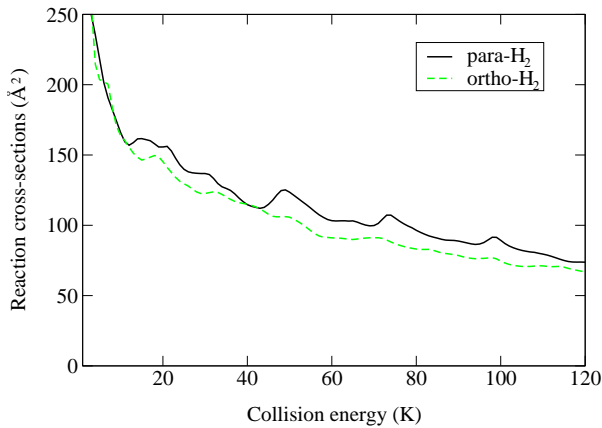,angle=-90,width=1.0\linewidth,clip=}
\caption{Quantum reaction cross-sections  for the collision of S($^1$D) with j=0 {\em p}-H$_2$ and j=1 {\em o}-H$_2$ at low kinetic energies. 
} \label{theoreticalcrosssections}
\end{center}
\end{figure}

\begin{figure}
\setlength{\unitlength}{8mm}
\begin{center}
\epsfig{file=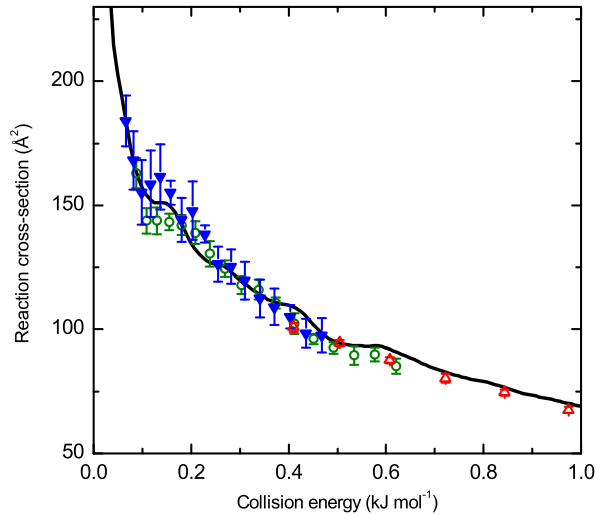,angle=0,width=1.0\linewidth,clip=}
\caption{Comparison of the excitation function obtained in the experiments by Costes and co-workers~\cite{Science} with our theoretical results. The latter were averaged using the ratio 3:1 of {\em o}-H$_2$  to {\em p}-H$_2$, and convoluted with a Gaussian distribution of collision energies to account for the velocity and angle spread in the experiment~\cite{Science}. The excitation function, in arbitrary units, was scaled in order to optimize the agreement with the theoretical data (see text). (Experimental data reproduced with the permission of the authors).} \label{experimentalcrosssections}
\end{center}
\end{figure}

  Quantum reactive cross-sections for collisions of S($^1$D) with j=0 {\em para-}hydrogen ({\em p}-H$_2$) and j=1 {\em ortho-}hydrogen ({\em o}-H$_2$), in the energy range 1$-$120 K, were calculated using the methodology described in the previous section.  In Fig.~\ref{theoreticalcrosssections} the obtained results are shown. A small dependence with the rotational state and some interesting oscillations, more apparent in the case of {\em p}-H$_2$, are relevant features. The cross-sections were averaged using the room-temperature ratio 3:1 of {\em o}-H$_2$ to {\em p}-H$_2$, so-called {\em normal-}hydrogen ({\em n}-H$_2$), which is expected to be valid in the conditions of the experiment by  Costes and co-workers~\cite{Science}. They were also convoluted with a Gaussian distribution of collision energies to account for the velocity and crossing angle spread in the experiment~\cite{Science}. In Fig.~\ref{experimentalcrosssections}, the experimental excitation functions (in arbitrary units) were scaled to minimize the sum of the squares of the differences between every experimental result and its corresponding theoretical value. Both sets show very similar decreasing trends as a function of the collision energy. 

\begin{figure}
\setlength{\unitlength}{8mm}
\begin{center}
\epsfig{file=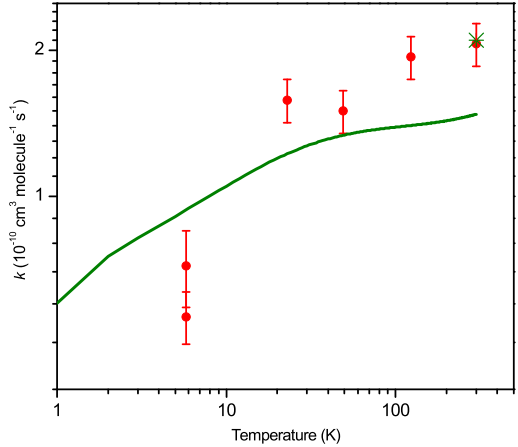,angle=0,width=0.9\linewidth,clip=}
\caption{Comparison of the experimental total removal rates (filled circles), obtained in the experiments by Sims and co-workers~\cite{nuestroFara, Science} with the thermally averaged reaction rate constant obtained in our calculations (dashed line). Again a ratio 3:1 of {\em o}-H$_2$  to {\em p}-H$_2$ was assumed. The 300 K experimental result of
Black and Jusinski\cite{Jusin} (marked with a cross) is also shown. (Experimental data reproduced with the permission of the authors).} \label{rates}
\end{center}
\end{figure}

Some additional calculations for particular collision energies in the range 150$-$1500K were performed in order to allow interpolations and the calculation of thermally-averaged rate constants to compare with the experimental results by Sims and co-workers~\cite{nuestroFara, Science}. Again, the room-temperature ratio 3:1 was considered. It was estimated that such proportion is not perturbed by the rapid cooling to 77 K and the subsequent adiabatic
expansion to 5.8 K in the experiment. By using a Maxwell-Boltzmann distribution the population of higher rotational states was estimated to be negligible for all the experimental temperatures except for the highest one, T=300K. 
Given the 
independency of the sets of rotational states with even and odd quantum number and the big rotational constant of H$_2$ ($\sim 85$K), only in the case of {\em p}-H$_2$ and for such high temperature the population of $j=2$ was found to be similar to  the population of the ground state. 
Accordingly, we calculated the cross-sections for $j=2$ at some particular collision
 energies 
and concluded a  difference of $\sim 5 \%$ between the cross-sections for $j=0$ and $j=2$. 
The contribution of  $j=2$ was not considered in the calculation of
 thermally averaged rate-constants and a value somewhat lower than the one in Fig.~\ref{rates} is expected for T=300K. 

 Interestingly, the theoretical rate constants approximate fairly well
the experimental ones, remaining below them  with the exception of the value for the lowest experimental temperature (5.8K). This temperature is also the most challenging from an experimental point of view. The 300 K experimental result of
Black and Jusinski\cite{Jusin} is also shown for comparison, in perfect agreement with the recent measurements. Let us remark that a multiplicative factor $1/5$ was included in the calculation of rate constants, as only one of the five asymptotically degenerate adiabatic surfaces is assumed to lead to the products in the considered energy range.

\section{Discussion}

\subsection{The significance of the long-range interactions} 
\begin{figure}
\setlength{\unitlength}{8mm}
\begin{center}
\epsfig{file=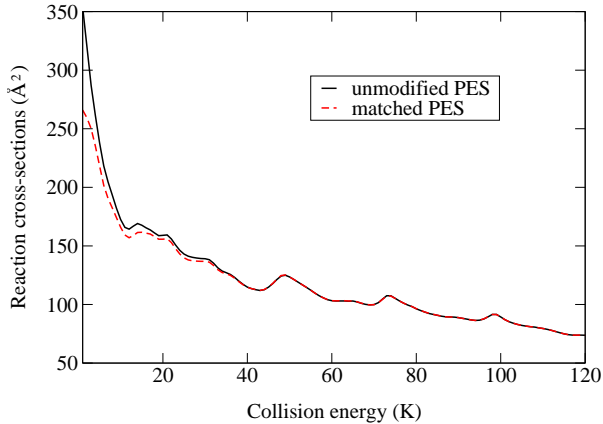,angle=-90,width=1.0\linewidth,clip=}
\caption{Sensitivity of the calculated cross-sections to the LR interations.
Cross-sections for the collision of S($^1$D) with {\em p}-H$_2$, obtained by using an uncorrected PES S1~\cite{Skodje02} are compared to the ones obtained by matching it with  a pure isotropic $\sim R^{-6}$ dispersion term, thus accounting for the expected LR effective behaviour.} \label{sinlong}
\end{center}
\end{figure}

In order to stress the need for accuracy in the description of LR effects while working at low collision energies, the sensitivity to such effects is shown in Fig.~\ref{sinlong}. Cross-sections for the collision with {\em p}-H$_2$, calculated using an uncorrected PES S1\cite{Skodje02}, are compared to the ones obtained by matching it with  a pure isotropic $\sim R^{-6}$ dispersion term
at LR (see Section~\ref{para}). As the figure shows, for energies below 20K the accurate inclusion of the LR interactions appears as essential. The use of S1 without corrections leads to differences of the order of 30$\%$ for energies around 1K.  

Note that the LR behaviour of an accurate {\em ab initio} surface for the ground electronic state contains a non-negligible $R^{-5}$ character in the title system. As argued above, such contribution  would effectively cancel while using a non-adiabatic approach (which considered the 5 asymptotically degenerate surfaces and all the non-adiabatic couplings). Nevertheless, it does not cancel while using an adiabatic approach (that being the reason to force a LR behaviour $R^{-6}$ in the calculations with {\em p}-H$_2$). The need for such artificial modifications reinforces the idea that for asymptotically degenerate systems an accurate treatment of the  LR non-adiabatic dynamics
seems essential. To the best of our knowledge,  non-adiabatic effects in low energy collisions have not been explored in depth up to the date.

\subsection{The electronic quenching}
In view of the results of Maiti {\em et al.}\cite{Maiti} and Tian-Shu {\em et al.}\cite{Tian} the electronic quenching is expected to play a role in the title collision. Since the experimental rate constants measured by Sims and co-workers~\cite{Science} probe the total removal of S($^1 D$) by {\em n}-H$_2$, they should comprise  the disappearance of S($^1 D$) due to the reactive process (formation of SH+H) or to the electronic quenching (relaxation to S($^3 P_{0,1.2}$)). The excitation functions measured by Costes and co-workers~\cite{Science} monitor only the first mechanism. As our theoretical calculations do not include the possibility of quenching some comments are pertinent in order to justify mutual comparisons.
What should we expect from the use of an electronically-uncoupled (EU) formalism to describe the dynamics when an electronically-coupled (EC) one would be more appropriate? A reasoning according to the statistical model of Rackham and Manolopoulos\cite{Rack0}, applied in the past to the 
system at thermal energies\cite{Rack1,Aoiztomas}, may help to clarify the limits and validity of our approach.  If the collisional process is assumed to be mediated by a long-lived complex (due to the deep well in the PES) we can distinguish the step of formation from the step of statistical decomposition of the complexes. Since the singlet-triplet crossing occurs inside the well we will consider the quenching as an alternative way of decomposition. Besides, we affirm that our calculation describes correctly the first step, the way the probability is captured. Regarding the second step, in a scenario where some complexes would decompose to give back the reactants (I), some to give reaction products (R), and the rest would suffer quenching (Q), our electronically-uncoupled (EU) calculation does not allow this latest possibility.
By making use of the principles of the model, and designing by $\sigma_I^{EC}$, $\sigma_R^{EC}$, $\sigma_Q^{EC}$ and $\sigma_I^{EU}$, $\sigma_R^{EU}$ the cross-sections for each possible outcome in an hypothetical EC (unprimmed) or in our EU (primmed) calculation, we will show that
  \begin{eqnarray}\label{eq:seccion}
\sigma_R^{EC} + \sigma_Q^{EC} > \sigma_R^{EU} > \sigma_R^{EC}
\label{secciones}
 \end{eqnarray}
 \noindent As a consequence, the theoretical rate constant calculated using $\sigma_R^{EU}$ constitutes a lower bound for the experimental total removal rate of S($^1 D$). The latter should be given by the sum of both reactive and quenching contributions in an exact EC calculation. 
 Besides, $\sigma_R^{EU}$ is itself an upper bound for the real reaction cross-section, monitored in arbitrary units by the measured excitation functions.  These statements should be valid even if $\sigma_R^{EU}$ is far from $\sigma_R^{EC}$. Below we will show how to reach these conclusions.

According to the statistical model~\cite{Rack0}, the fraction of complexes which decompose into a particular  channel is related to the {\em capture} probability of forming the complex starting from that particular channel. 
Let us denote with $A^{J}_n (E)$, $B^{J}_{n'}(E)$ and $C^{J}_{n''}(E)$  the energy dependent capture probabilities starting from reagent, 
product and quenching sides, corresponding to particular channels $n,n',n''$, respectively.  $J$ labels the total angular momentum.
 Let us define the sum in open channels of the capture probabilities associated to each possible outcome, 
$A^J(E)=\sum  A^{J}_n (E) $, $B^J(E)=\sum B^{J}_{n'}(E)$ and $C^J(E)=\sum C^{J}_{n''}(E) $. Following the prescriptions of the model, the reaction
 probability corresponding to the reagents colliding in channel $n$,  $P^J_{R~n}(E)$, is given by $ P^{J \ EU}_{R~n}(E)= A^{J}_n (E)B^J(E) /(A^J(E)+B^J(E))$  in our EU calculation and by $ P^{J \ EC}_{R~n}(E)= A^{J}_n (E)B^J(E) /(A^J(E)+B^J(E)+C^J(E))$ in an hypothetical EC calculation. Interestingly, the former is larger. Now, essentially by summing over $J$, one obtains cross-sections and it follows that $\sigma_{R}^{EU}> \sigma_{R}^{EC}$, which is the second inequality in expression~\ref{secciones}. A similar reasoning leads from the probability of total removal, $P^{J \ EC}_{R~n}(E)+P^{J \ EC}_{Q~n}(E)= A^{J}_n (E)(B^J(E)+C^J(E)) /(A^J(E)+B^J(E)+C^J(E))$,  to the first inequality. In agreement with it, the thermally averaged theoretical rates in Fig.~\ref{rates} remain below the experimental ones, 
with the exception of the value for the lowest temperature. 
\subsection{The Langevin-type capture dynamics}
Both the theoretical and the experimental
excitation functions,  in Fig.~\ref{experimentalcrosssections}, show no threshold,
consequence of the attractive character without barrier of the reaction on the ground electronic state.
Remarkably, both follow very similar trends as a function of the kinetic energy, in spite of the absence of quenching channels in the theoretical calculations. As a first guess, such proportionality would result from the {\em real} reaction probabilities following the entrance channel capture probabilities, $A^J_n(E)$, given that the capture is well described in our approximate calculation. Moreover, if our guess were correct the experimental reaction cross-sections should have a Langevin-capture character\cite{Lange} (see note~\footnote{The Langevin capture model assumes that the system is always captured in the complex when the kinetic energy overcomes the centrifugal barrier and,  once the complex is formed, that it always leads to products. The former means that capture probabilities, $A^J_n(E)$, change from zero to one when overcoming the centrifugal barrier. The latter identifies capture probabilities with reaction probabilities, thus assuming the upper limit for the proportion of complexes which decompose into products.}).

Such qualitative reasoning can be put in solid grounds by making use again of the statistical model. 
Given the exothermicity of both the reactive and quenching processes, many internal states are open even at low collision energies in the corresponding arrangements. In particular, $B^J(E)$ may be approximated by the number of open channels, thus being much larger than $A^J(E)$. The value of $C^J(E)$ is more difficult to estimate as the capture from the quenching side requires a non-adiabatic transition, of unknown probability, to occur. 
Regardless, it is resonable to accept that
 $B^J(E)$, and thus $B^J(E)+C^J(E)$, are much bigger than $ A^J(E)$ in the energy and  angular momentum ranges which contribute, thus following that most of the complexes are expected to form SH+H or to suffer quenching. Besides, such figures are expected to be essentially constant, $B^J(E) \sim B$, $B^J(E)+C^J(E) \sim B+C $. 
 Accordingly, in we neglect the contribution of
 $ A^{J} (E)$ in the denominator of $ P^{J \ EU}_{R~n}(E)$ (see above) we find a result which is essentially equal to a reactant capture probability, $A^{J}_n (E)$.
Proceeding similarly we find that  $ P^{J \ EC}_{R~n}(E)$ is a constant portion of the capture probability, $ P^{J \ EC}_{R~n}(E) \sim (B /(B+C)) A^{J}_n (E)$, and thus proportional to $ P^{J \ EU}_{R~n}(E)$.
Finally, the sum $ P^{J \ EC}_{R~n}(E)+P^{J \ EC}_{Q~n}(E)$ is again equal to the capture probability $A^{J}_n (E)$ and therefore to $ P^{J \ EU}_{R~n}(E)$.  By summing over $J$ to obtain cross-sections, the previous results are found equivalent to the following statements: i) $\sigma_{R}^{EU}$  is essentially a capture cross-section which accounts for the total number of complexes formed in the collision, ii) $\sigma_{R}^{EU}$ should be found {\em proportional} to $\sigma_{R}^{EC}$, and iii) $\sigma_{R}^{EC}+ \sigma_{Q}^{EC} $ and  $\sigma_{R}^{EU}$  must be similar to each other. We will check their validity below.

\begin{figure}
\setlength{\unitlength}{8mm}
\begin{center}
\epsfig{file=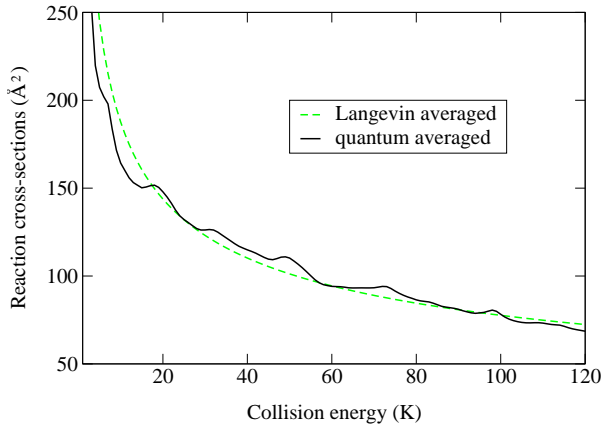,angle=-90,width=1.0\linewidth,clip=}
\caption{In the figure, the theoretical cross-section averaged using the  3:1 proportion of {\em o}-H$_2$  to {\em p}-H$_2$ present in the experiment by Costes and co-workers~\cite{Science} is compared with the averaged analytical Langevin estimate\cite{Lange} (see text).
} \label{Langevv}
\end{center}
\end{figure}

Regarding the first statement,  the quantum cross-section averaged for {\em n}-H$_2$ is compared in Fig.~\ref{Langevv} with an averaged Langevin capture estimate and a good agreement is found. A capture dispersion potential $-C_6/R^6$ was  assumed for {\em p}-H$_2$, what leads to an analytical Langevin cross-section given by $\sigma_{Lan}^6(E)= 3 \pi \left( C_6 / 4E \right)^{1/3}$. A QQ electrostatic potential,  $-C_5/R^5$, and the corresponding cross-section,  $\sigma_{Lan}^5(E)= 5 \pi \left(C_5 / 6 \sqrt{3} E \right)^{2/5}$, were assumed for {\em o}-H$_2$ instead. The values $C_6=C_{6,k=0}^{0,0}=38.9$ a.u.  and $C_5=4.08~\langle \hat{Q}_{20}\rangle_{S} \langle \hat{Q}_{20}\rangle_{H_2}$  were respectively taken. The former is the lowest isotropic dispersion coefficient and the latter results from an angular average of the ground eigenvalue of the QQ contribution in Eq.~\ref{LR}. Double weight is assigned to the perpendicular approach, $C_5(\theta=90)=\sqrt{39}/2~\langle \hat{Q}_{20}\rangle_{S} \langle \hat{Q}_{20}\rangle_{H_2}$, with respect to the linear one, $C_5(\theta=0)=6~\langle \hat{Q}_{20}\rangle_{S} \langle \hat{Q}_{20}\rangle_{H_2}$, the factor  4.08 resulting from such weighted average. The second statement explains the similar trends found between theoretical and experimental excitation functions, in spite of the non-inclusion of quenching in the calculations. Finally, the third statement justifies the similarity of the experimental and theoretical rates, shown in Fig.~\ref{rates}.
 The bigger discrepancy between the experimental and theoretical rate coefficients at 5.8 K, the latter being larger, may be related with the neglected non-adiabatic couplings between the three singlet A' states which correlate with the reactants.  In fact, the discrepancy can be interpreted as the theoretical simulations capturing more probability in the complex than they should, and neglecting the couplings to repulsive states could be a reason for such an effect. The effect of such couplings should increase dramatically when decreasing the kinetic energy, while the contribution from the neglected singlet-triplet couplings is expected to be less energy dependent. Regardless, such low kinetic energies should be more sensitive to possible inaccuracies of the considered PES.

 To outline this section, let us remark that our theoretical cross-sections are essentially capture cross-sections and they must account well for the total number of complexes formed in the collision. The number of complexes which eventually react is found a constant portion of the total number. This explains the experimental reaction cross-sections being proportional to the theoretical cross-sections. Finally, as the experimental total removal rates include the sum of reaction and quenching, they are monitoring again the total number of complexes, so they are well approximated by our theoretical results. 

\subsection{The structures in the cross-sections}
\begin{figure}
\setlength{\unitlength}{8mm}
\begin{center}
\epsfig{file=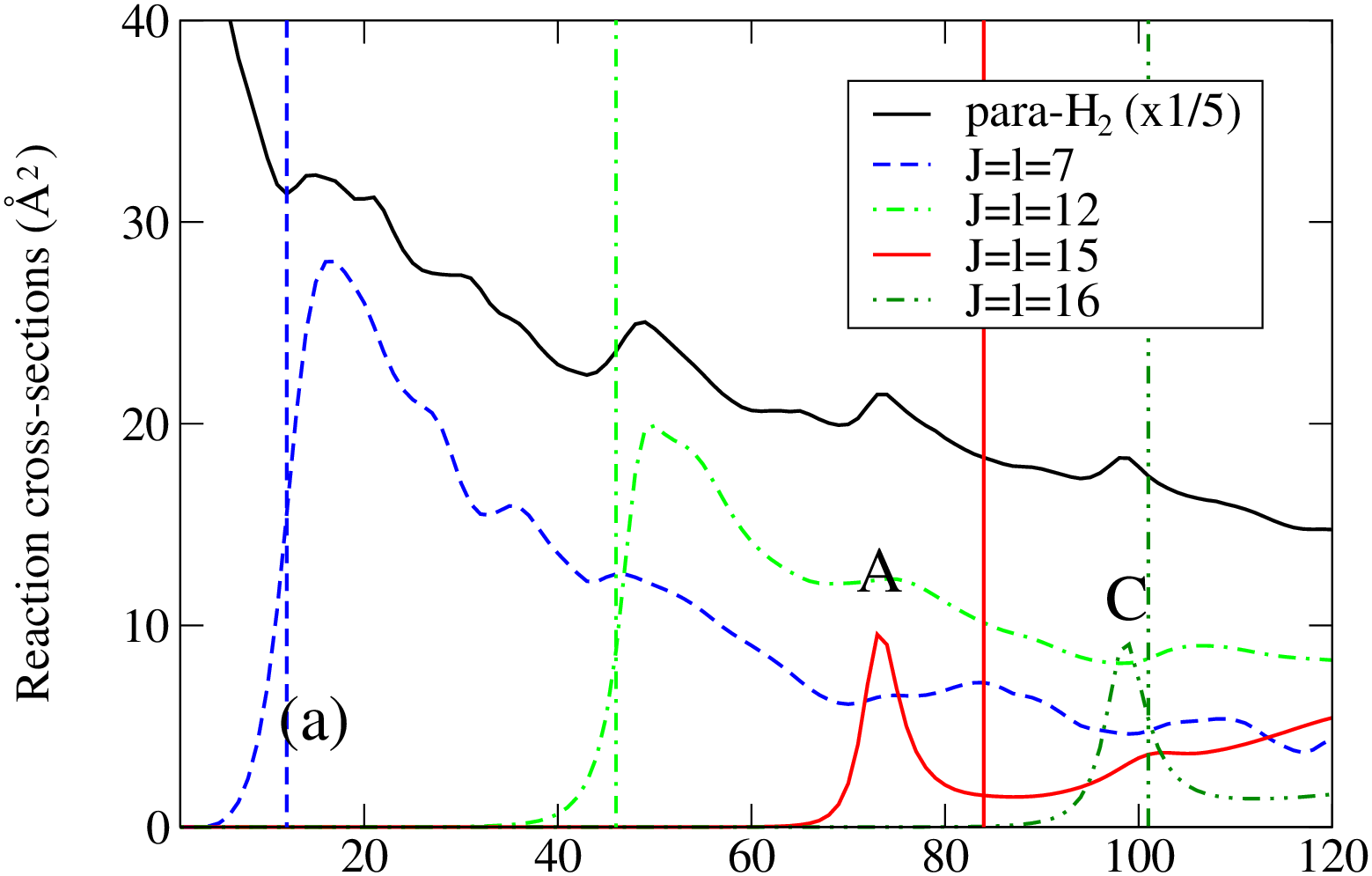,angle=-90,width=1.0\linewidth,clip=}
\epsfig{file=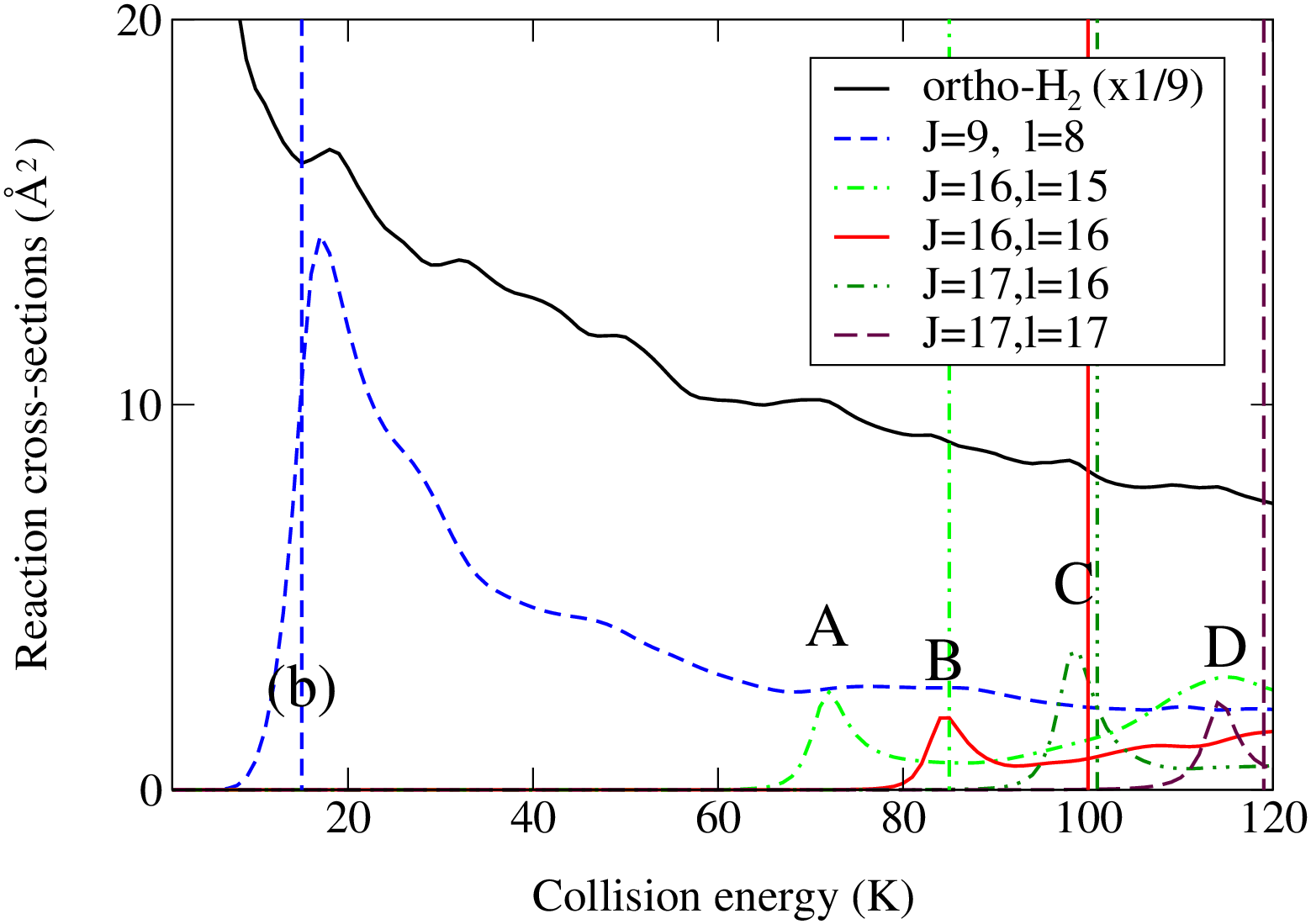,angle=-90,width=1.0\linewidth,clip=}
\caption{In the figure, some oscillation observed in the quantum reaction cross-sections are attributed to particular partial waves
and entrance channels. 
Panel (a) corresponds to j=0 {\em p}-H$_2$ and  
panel (b)  
to j=1 {\em o}-H$_2$. Peaks
associated to $J=l=7$ and  $J=l=12$ in the case of {\em p}-H$_2$ 
, and  $J=9,l=8$ in the case of {\em o}-H$_2$
occur at energies just above the corresponding centrifugal barriers (shown as vertical lines in the same color or type of line).
In contrast, peaks A, B, C and D occur below barrier (see text).
}\label{oscillations}
\end{center}
\end{figure}

The theoretical cross-sections show interesting oscillations  superimposed to
the overall Langevin behaviour (Fig.~\ref{oscillations}).
They can be observed for both  {\em p}-H$_2$ and {\em o}-H$_2$, although their presence seems to be washed out by the sum in partial waves in the latter case. In a {\em naive} approach, zero-order resonances present in the case of {\em p}-H$_2$ would split in the case of {\em o}-H$_2$ due to the different values of $\Omega$. Their abundance and mutual overlap would justify their being more difficult to observe in the latter case. 

 \begin{figure}
\setlength{\unitlength}{8mm}
\begin{center}
\epsfig{file=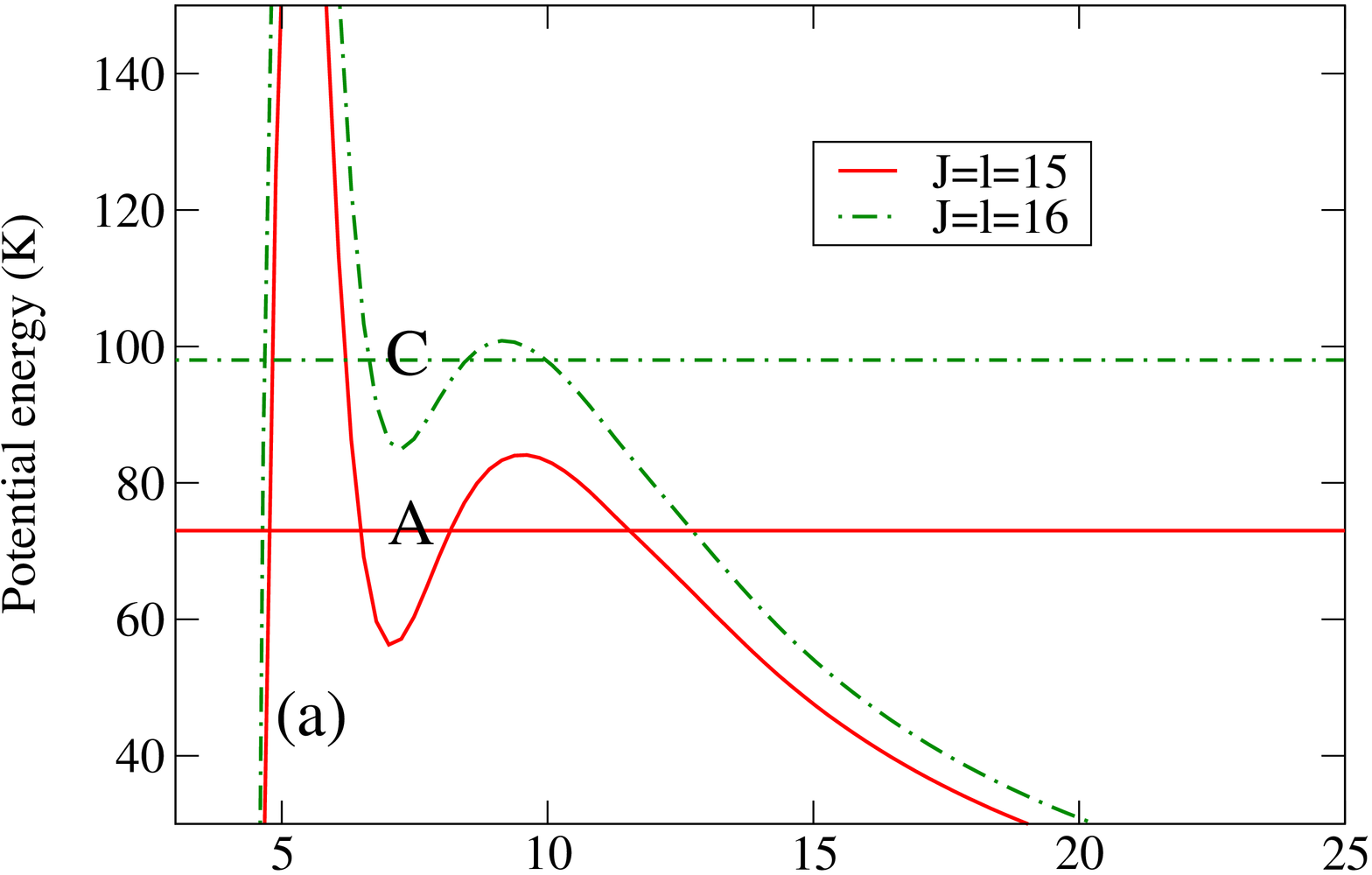,angle=-90,width=1.0\linewidth,clip=}
\epsfig{file=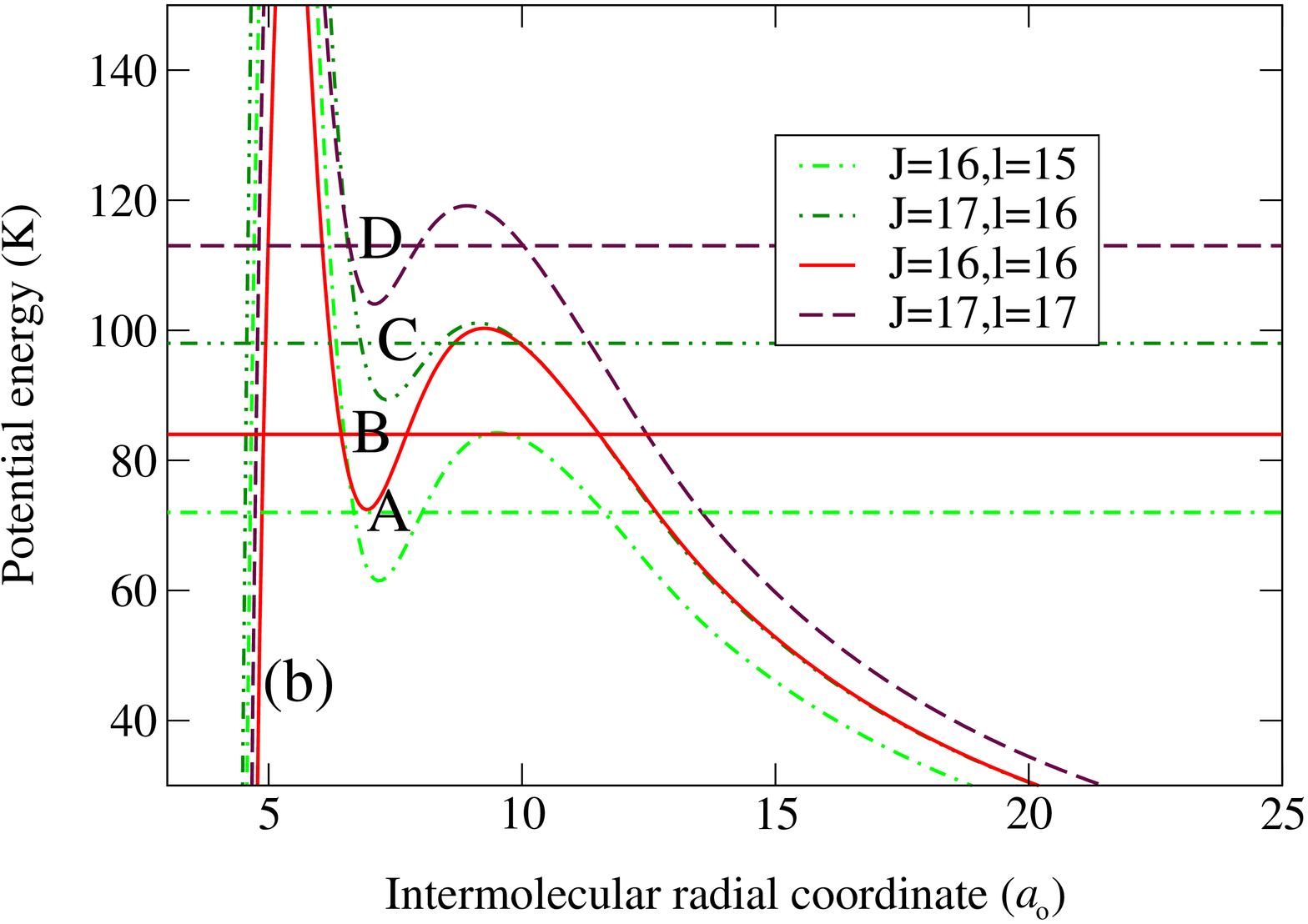,angle=-90,width=1.0\linewidth,clip=}
\caption{In the figure, the effective potentials, $\langle  V \rangle_{l,l}(R)+ l(l+1) \hbar^2 /2 \mu R^2$, of the channels associated to peaks  A, B, C and D (see Fig.~\ref{oscillations}) are shown to provide the necessary trapping to support quasibound states in the form of shape resonances. Panel (a) corresponds to j=0 {\em p}-H$_2$ and 
panel (b) to j=1 {\em o}-H$_2$.  
The horizontal lines mark the location in energy of the peaks. Note the similarity of the effective potentials corresponding to the pair of channels ($J=16, l=15$), ($J=17,l=16$) of {\em o}-H$_2$ to the ones corresponding to $J=l=15$, $J=l=16$ of {\em p}-H$_2$.
.} \label{effective}
\end{center}
\end{figure}
 
 The observed structures, which can be attributed to particular partial waves and channels have a double origin: some of them  display the effective 
 opening of a partial-wave $l$; others seem to be due to the presence of shape resonances. 
Regarding the former, the Wigner threshold laws provide a behaviour for the $l$ cross-section below the centrifugal barrier given by $E^{l-1/2}$. 
 If, after the barrier were overcome, the reaction probability reached quickly the limit value of one, 
the behaviour should follow a $1/k^2$ dependence, that is $\sim E^{-1}$\cite{Quem05,SolPRL,QueB}. This would result as a ``bump'' in the cross-sections. 
A bump of this type should happen at energies over the height of the centrifugal barrier. On the contrary, oscillations for high partial waves steming from shape resonances would occur at energies below the centrifugal barriers, the associated effective potentials providing some kind of trapping.

 Some structures  have been selected and  they are shown in Fig.~\ref{oscillations} to be associated to particular partial waves and channels, $\phi_{vjl}^{JM}$. The corresponding centrifugal barriers are also displayed as vertical lines. 
They were calculated using effective potentials, $\langle  V \rangle_{ l, l} (R)+l(l+1)\hbar^2/ 2 \mu R^2$, where the average $\langle  V \rangle_{ l, l} (R)$ was defined in Eq.~\ref{camio}. 
Panel (a) corresponds to j=0 {\em p}-H$_2$ and panel (b) to j=1 {\em o}-H$_2$.
Given the shape and location (above the centrifugal barrier) of the peaks associated to {\em p}-H$_2$, with entrance channels  $J=l=7$ 
and $J=l=12$, and to {\em o}-H$_2$ with entrance channel $J=9,l=8$, we conclude that their origin is the opening of a new partial wave. 
In contrast, the peaks labelled A and C for {\em p}-H$_2$, associated to the $J=l=15$ and $J=l=16$ contributions, are found below barrier. The same happens for the  peaks  A, B, C and D in the lower pannel, associated to the entrance in channels ($J=16, l=15$),  ($J=16, l=16$),  ($J=17, l=16$)  and  ($J=17, l=17$), respectively. As we will explain below, we attribute them to the presence of shape resonances. Note that peaks A and C are found at similar energies for {\em o}-H$_2$ and {\em p}-H$_2$. 

 The presence of multiple resonances in the reaction probability of the title system for zero total angular momentum was reported in the past~\cite{Banares02}. They manifest the indirect mechanism through the formation of long-lived complexes. If the peaks selected in the current work were resonances, the associated trapping should happen in an intermediate range of radial distances. 
On the one hand, they appeared in calculations where the LR interaction was not well accounted for (see Fig.~\ref{sinlong}), so they cannot be located at very large distances. On the other, their location at very short range would hamper, due to the high anisotropy of SR chemical forces, any rotational adiabaticity which may lead to similar effective potentials, and thus to the similar structures observed, for j=0 and j=1. 
In good agreement with such qualitative reasoning, the PES shows a shallow well in the entrance valley at distances of $\sim 7$ a.u.~\cite{Aoiztomas}. With a depth of $\sim 500$K for collinear approaches, and appearing as a small
 secondary minimum opening the access to the big well for insertion paths, this complex is the probable origin of many of the structures in the cross-sections.  As a result of such well, the effective potentials of the channels associated to the peaks show a double maximum structure (see Fig.~\ref{effective}). The local minimum inbetween may support shape resonances at precisely the energies where the structures appear. Remarkably, the effective potentials corresponding to channels ($J=16, l=15$) and  ($J=17,l=16$)  of {\em o}-H$_2$ are respectively very similar to the ones corresponding to $J=l=15$ and  $J=l=16$ of {\em p}-H$_2$. This justifies the appearance of peaks A and C at similar energies in both cases.
The approaching partners would be trapped in this preliminar well until they tunneled into the big one, thus finding the way to reaction, or they tunneled backwards, thus giving the reagents back. The quasibound states seem to prefer  ``prereaction''though, increasing reactivity at the associated energies. Note finally that Fano-type profiles, which indicate quick changes in the phase of the elastic S-matrix elements, are found in the elastic cross-sections at similar energies, thus reinforcing the hypothesis about the nature of the peaks as resonances.

Let us remark that the role of van der Waals wells in reactive collisions at very low energy was recently stressed~\cite{Balak}. As stated, the supported quasibound states may lead to a temporal trapping of the system at short distances, and to an increase of tunneling through the barrier. Such effects were mainly discussed  for tunneling-dominated reactions like Li+HF\cite{WeckLihf} and Cl+HD~\cite{Balak}, where the reaction would be closed otherwise. In addition to the considered shape resonances, Feshbach-type ones may occur for zero angular momentum and can dramatically modifying ultracold reaction cross-sections and zero-temperature rate constants~\cite{bodoG}.

\section{Conclusions}
In this work, the S($^1$D) + H$_2$ collision was analyzed in the kinetic energy range 1$-$120K. Within an adiabatic treatment, working on the lowest surface which correlates with the reagents, cross-sections and thermally averaged rate constant were obtained using the  quantum reactive scattering method developed by Launay {\em et al.}~\cite{Launayfirst}. The resulting picture of the process is the following: the behavior can be understood on the average in terms of a classical Langevin (capture) model where centrifugal barriers determine the amount of reactive flux which reaches the barrierless transition state; additionally, the structure of the van der Waals well provides temporal trapping at short distances, thus helping the system to find its way to reaction for high partial waves. Comparison with landmark experimental data obtained using the  crossed molecular beam machine with variable beam intersection-angle~\cite{Costes} (Bordeaux) and the CRESU technique~\cite{Canosa,nuestroFara} (Rennes) shows a fairly good agreement.  Accurate electronic calculations of the LR interactions were required in order to propertly describe collisions at such low kinetic energy. In particular, the anisotropic QQ term of the interaction appears to be important to describe the collision with {\em o}-H$_2$, although it does not contribute in the case of {\em p}-H$_2$. 

 Previous theoretical works at higher energies indicated that the
electronic quenching process, and thus the intersystem crossing, may play a significant role in the total
removal of S($^1$D)\cite{Maiti,Tian}. Nevertheless, a good agreement of our theoretical and the experimental total removal rates has been found in spite of the adiabatic treatment of the collision. Moreover, the reescaled 
experimental excitation functions follow very well the
theoretical ones. An small contribution of the quenching must not be concluded though. The current theory accounts well for the total number of complexes formed in the collision, this amount being what our theoretical rates and cross-sections essentially monitor. Now, as most of the complexes decompose to give either reaction or quenching, and experimental removal rates sum up both contributions, a good agreement of theoretical and removal rates follows. Besides, the ratio of reaction to quenching is not expected to significantly change in the 
small considered energy range. The amount of complexes which 'react' at one particular energy appears then a constant fraction of the total number of complexes captured at such energy. A proportionality of the 'real' reaction cross-section to our theoretical cross-sections automatically follows, and therefore the good agreement with the experimental excitation functions. It must be noted that both discussed agreements seem to follow independently to the proportion of quenching and, in our opinion, the question remains open. Future measurements of the branching ratio reaction/quenching would be thus desirable.

 Interestingly, the theoretical cross-sections showed oscillating features. 
 We conclude that some of them are associated to the presence of a small van der Waals well  in the entrance valley of the {\em ab initio} PES S1~\cite{Skodje02}. Far from being an artifact of the fit, this well seems to be real, appearing in some {\em ab initio} calculations we have performed to verify its existence. Nevertheless, we have found a much smaller depth. Although the current experimental results do not allow conclusive confirmation of the presence and nature of the oscillating structures, the isolation of the contribution of {\em p}-H$_2$ in the cross-beam experiment or the  complementary measurement of differential cross-sections could help to asess their existence and origin~\cite{Xueming}. If they were resonances, they would be very sensitive to the potential energy surface and they would give us important information about it. Note that 
due to the high degree of averaging into orientations, energies and directions of approach, direct observation of resonant structures in reaction cross-sections at thermal energies has proven ellusive\cite{Schatz}. Up to the date it was only successful\cite{Skodje} in the F+HD system, and thanks to the concurrence of a set of favorable dynamical conditions. The current experimental access to the low-temperature regime, charaterized by small partial-wave averaging, may lead to a different scenario. 

Finally, our work reinforces the idea that an accurate treatment of non-adiabatic dynamics at LR for asymptotically degenerate systems 
seems fundamental at low energies. Previous studies at thermal energies stressed the redistribution of asymptotic flux onto the different PES which may result from such degeneracy while the reagents approach\cite{Druk}. The role  of such non-adiabatic connections at low energies has not been explored yet. A better description of the title process including non-adiabatic effects is in progress.

\acknowledgments
This research is supported by the Agence Nationale de la Recherche
under contract ANR-BLAN-2006-0247, {\em Cold Reactions between Neutral Species}, which
provided fellowships for M. L. and F. D. The authors acknowledge F. J. Aoiz for his very useful comments on the manuscript.

\section{Appendix : Long-range electrostatic and dispersion energies}
\label{appendix}

We consider below the case of LR interactions between an open-shell atom, $A$, and a closed-shell diatom, $B$. To account for the $n$-fold degeneracy of the atom, we define a set of asymptotically degenerate diabatic states, taken as a product of the atomic $\ket{L\lambda}^{(A)}$ and diatomic $\ket{0}^{(B)}$ unperturbed electronic states.  The quantum number $\lambda=-L,\ldots,L$ is the projection of the atomic orbital angular momentum ${\bf L}$ along the Body-Fixed (BF) $z$-axis, chosen along the intermolecular vector ${\bf R}$, and the projection $\lambda=0$ of the diatomic orbital angular momentum refers to the diatom axis ${\bf r}$, chosen to be in the BF $xz$-plane. 

The multipolar expansion of the first-order, ${\hat H}_{el}^{(1)}$, and second-order, ${\hat H}_{el}^{(2)}$, perturbation operators depend on multipole moments ${\hat Q}^{(A)}_{l_1m_1}$ and ${\hat Q}^{(B)}_{l_2m_2}$ that are defined in frames centered on $A$ and $B$, with axes parallel to a global Space-Fixed (SF) frame. To treat the atom-diatom case, we introduce multipole moments ${\hat Q}^{(B)}_{l_2\mu_2}$ that are defined relative to a local reference frame on $B$, 

\begin{equation}
{\hat Q}^{(B)}_{l_2m_2}=\sum_{\mu_2}{\hat Q}^{(B)}_{l_2\mu_2}\left[D^{l_2}_{m_2\mu_2}({\hat\Omega}_B)\right]^{\ast},
\end{equation}

\noindent where ${\hat\Omega}_B$ collects the polar angles of the diatom axis ${\bf r}$ relative to the SF frame, and 
$D^{l_2}_{m_2\mu_2}({\hat\Omega}_B)$ is a Wigner rotation matrix. By rotating the SF frame such as the $z$-axis coincides with ${\bf R}$, and ${\bf r}$ lies in the $xz$-plane, the Wigner rotation matrix reduces to $\left[D^{l_2}_{m_2\mu_2}({\hat\Omega}_B)\right]^{\ast}=d^{l_2}_{m_2\mu_2}(\theta)$, with $\theta$ being the angle between ${\bf R}$ and ${\bf r}$.

 Following previous works on open-shell interacting systems~\cite{bussery:08,groenenboom:06,spelsberg:99,zeimen:03}, the matrix elements of ${\hat H}^{(1)}_{el}$ in the $\ket{L\lambda}^{(A)}\ket{0}^{(B)}$ basis set describe the electrostatic energies and, according to the chosen orientation for the global frame, they can be written as:

\begin{equation}
\label{elec}
V^{\rm elec}_{\lambda,\lambda'}(R,\theta)=\sum_{l_1l_2m}\frac{1}{R^{l_1+l_2+1}}\; V^{\lambda\lambda'}_{l_1l_2m}\; C_{l_2,-m}(\theta,0)
\end{equation}

\noindent where the angular functions $C_{l,m}(\theta,\phi)$ are normalized spherical harmonics, and the electrostatic interaction coefficients are given by:

\begin{eqnarray}
\label{coeffelec}
V^{\lambda\lambda'}_{l_1l_2m} & = & (-1)^{l_2}\binom{2L_{12}}{2l_1}^{1/2}\left<l_1ml_2-m\right.\ket{L_{12}0} 
\\ \nonumber
&&\times \bra{L\lambda}\hat{Q}^{(A)}_{l_1m}\ket{L\lambda'}\;\bra{0}\hat{Q}^{(B)}_{l_20}\ket{0}
\end{eqnarray}

\noindent where $L_{12}=l_1+l_2$. From the Wigner-Eckart theorem, it follows that the multipole moments matrix elements $\bra{L\lambda}\hat{Q}^{(A)}_{l_1m}\ket{L\lambda'}$ and $\bra{0}\hat{Q}^{(B)}_{l_2\mu_2}\ket{0}$ vanish unless $m=\lambda-\lambda'$ and $\mu_2=0$, respectively. For the sake of clarity, the electrostatic interaction coefficient $V_{l_1l_2m}^{\lambda\lambda'}$ will be substituted hereafter by the label $C^{\lambda\lambda'}_{l_1+l_2+1}$, since it is commonly used in the context of LR interactions.

The matrix elements of ${\hat H}^{(2)}_{el}$ in the $\ket{L\lambda}^{(A)}\ket{0}^{(B)}$ basis set describe the dispersion energies, and write as~\cite{bussery:08,groenenboom:06,spelsberg:99}:

\begin{equation}
\label{disp}
V^{\rm disp}_{\lambda,\lambda'}(R,\theta)=-\sum_{l_1l_2l'_1l'_2}\frac{1}{R^{L_{12}+L'_{12}+2}}\sum_{k_2M}
V^{\lambda\lambda'}_{l_1l'_1l_2l'_2;k_2M}\; C_{k_2,-M}(\theta,0)
\end{equation}

\noindent where the dispersion interaction coefficients are given by:

\begin{equation}
\label{coeffdisp}
V^{\lambda\lambda'}_{l_1l'_1l_2l'_2;k_2M} = \sum_{k_1k} f^{k_1k_2k}_{l_1l'_1l_2l'_2}\; 
\left<k_1Mk_2-M\right.\ket{k0}\; {\cal X}^{\lambda\lambda'}_{(l_1l'_1)k_1(l_2l'_2)k_2}
\end{equation}

\noindent The angular coupling coefficient $f^{k_1k_2k}_{l_1l'_1l_2l'_2}$ is defined as in Eq. 21 of Ref.~\cite{groenenboom:06}, and the quantity ${\cal X}^{\lambda\lambda'}_{(l_1l'_1)k_1(l_2l'_2)k_2}$ is a coupled form of Casimir-Polder integral:

\begin{equation}
\label{casimir}
{\cal X}^{\lambda\lambda'}_{(l_1l'_1)k_1(l_2l'_2)k_2}=\frac{1}{2\pi}\int_0^{\infty}\  ^{\lambda\lambda'}\alpha^{(A)}_{(l_1l'_1)k_1M}(i\omega)\  ^{00}\alpha^{(B)}_{(l_2l'_2)k_20}(i\omega)\; d\omega .
\end{equation}

\noindent From the Wigner-Eckart theorem, it follows that the coupled dynamic polarizabilities $^{\lambda\lambda'}\alpha^{(A)}_{(l_1l'_1)k_1M}$ and $^{00}\alpha^{(B)}_{(l_2l'_2)k_2Q_2}$ vanish unless $M=\lambda-\lambda'$ and $Q_2=0$, respectively. Furthermore, in the particular case of dipole-dipole polarizabilities ($l_2=l'_2=1$) and a diatomic state ($|\lambda|=|\lambda'|$), the matrix elements $^{\lambda\lambda'}\alpha^{(B)}_{(l_2l_2)k_2Q_2}$ vanish unless $Q_2=\lambda-\lambda'$ and $k_2$ is even. Again, for the sake of clarity, we substitute hereafter the dispersion interaction coefficient $V^{\lambda\lambda'}_{l_1l'_1l_2l'_2;k_2M}$ by the more commonly used label $C^{\lambda\lambda'}_{l_1+l_2+l'_1+l'_2+2,k_2}$.

In present work, we have considered the quadrupole-quadrupole interactions ($l_1=l_2=2$) between an open-shell atom $A$ and a closed-shell diatom $B$. Hence, the quadrupole moments $\bra{L\lambda}\hat{Q}^{(A)}_{2(\lambda-\lambda')}\ket{L\lambda'}$ and $\bra{0}\hat{Q}^{(B)}_{20}\ket{0}$ are the necessary ingredients to compute the LR electrostatic coefficient $C^{\lambda\lambda'}_{5}$ of Eq.~\ref{coeffelec}. Besides, we have considered the dipole-induced dipole-induced interactions ($l_1=l'_1=l_2=l'_2=1$) between the two species. In such a case, the LR dispersion coefficient $C^{\lambda\lambda'}_{6,k_2}$ of Eq.~\ref{coeffdisp} vanish unless $k_1$, $k_2$ and $k$ take even values, and the necessary ingredients are the coupled dipole-dipole dynamic polarizabilities $^{\lambda\lambda'}\alpha^{(A)}_{(11)k_1(\lambda-\lambda')}$ and $^{00}\alpha^{(B)}_{(11)k_20}$, with $k_1=0,2$ and $k_2=0,2$. Notice that the matrix elements are here defined in terms of spherical components of the $\hat{Q}_{lm}$ multipole moment and $\hat{\alpha}_{lml'm'}$ dynamic polarizability operators, together with a basis set of (complex) signed-$\lambda$ electronic states. Nonetheless, we can benefit from the Wigner-Eckart theorem to consider only the diagonal matrix elements, which have one-to-one correspondence with the quantities derived from quantum chemistry calculations.

\bibliography{Majorana2}

\begin{thebibliography}{75}
\expandafter\ifx\csname natexlab\endcsname\relax\def\natexlab#1{#1}\fi
\expandafter\ifx\csname bibnamefont\endcsname\relax
  \def\bibnamefont#1{#1}\fi
\expandafter\ifx\csname bibfnamefont\endcsname\relax
  \def\bibfnamefont#1{#1}\fi
\expandafter\ifx\csname citenamefont\endcsname\relax
  \def\citenamefont#1{#1}\fi
\expandafter\ifx\csname url\endcsname\relax
  \def\url#1{\texttt{#1}}\fi
\expandafter\ifx\csname urlprefix\endcsname\relax\def\urlprefix{URL }\fi
\providecommand{\bibinfo}[2]{#2}
\providecommand{\eprint}[2][]{\url{#2}}

\bibitem[{\citenamefont{Qu{\'{e}}m{\'{e}}ner
  et~al.}(2009)\citenamefont{Qu{\'{e}}m{\'{e}}ner, Balakrishnan, and
  Dalgarno}}]{coldcol}
\bibinfo{author}{\bibfnamefont{G.}~\bibnamefont{Qu{\'{e}}m{\'{e}}ner}},
  \bibinfo{author}{\bibfnamefont{N.}~\bibnamefont{Balakrishnan}},
  \bibnamefont{and} \bibinfo{author}{\bibfnamefont{A.}~\bibnamefont{Dalgarno}},
  \emph{\bibinfo{title}{in {\sl Cold Molecules: Theory, Experiment,
  Applications}}} (\bibinfo{publisher}{CRC Press}, \bibinfo{year}{2009}).

\bibitem[{\citenamefont{Julienne}(2009)}]{Fara}
\bibinfo{author}{\bibfnamefont{P.}~\bibnamefont{Julienne}},
  \bibinfo{journal}{Faraday Discuss.} \textbf{\bibinfo{volume}{142}},
  \bibinfo{pages}{361} (\bibinfo{year}{2009}).

\bibitem[{\citenamefont{Bell and Softley}(2009)}]{Bell}
\bibinfo{author}{\bibfnamefont{M.~T.} \bibnamefont{Bell}} \bibnamefont{and}
  \bibinfo{author}{\bibfnamefont{T.~P.} \bibnamefont{Softley}},
  \bibinfo{journal}{Mol. Phys.} \textbf{\bibinfo{volume}{107}},
  \bibinfo{pages}{99} (\bibinfo{year}{2009}).

\bibitem[{\citenamefont{Hutson and Sold{\'{a}}n}(2006)}]{Hutsonrev}
\bibinfo{author}{\bibfnamefont{J.~M.} \bibnamefont{Hutson}} \bibnamefont{and}
  \bibinfo{author}{\bibfnamefont{P.}~\bibnamefont{Sold{\'{a}}n}},
  \bibinfo{journal}{Int. Rev. Phys. Chem.} \textbf{\bibinfo{volume}{25}},
  \bibinfo{pages}{497} (\bibinfo{year}{2006}).

\bibitem[{\citenamefont{Balakrishnan and Dalgarno}(2001)}]{UltraD}
\bibinfo{author}{\bibfnamefont{N.}~\bibnamefont{Balakrishnan}}
  \bibnamefont{and} \bibinfo{author}{\bibfnamefont{A.}~\bibnamefont{Dalgarno}},
  \bibinfo{journal}{Chem. Phys. Lett.} \textbf{\bibinfo{volume}{341}},
  \bibinfo{pages}{652} (\bibinfo{year}{2001}).

\bibitem[{\citenamefont{Bodo et~al.}(2004)\citenamefont{Bodo, Gianturco,
  Balakrisnan, and Dalgarno}}]{elramsa}
\bibinfo{author}{\bibfnamefont{E.}~\bibnamefont{Bodo}},
  \bibinfo{author}{\bibfnamefont{F.~A.} \bibnamefont{Gianturco}},
  \bibinfo{author}{\bibfnamefont{N.}~\bibnamefont{Balakrisnan}},
  \bibnamefont{and} \bibinfo{author}{\bibfnamefont{A.}~\bibnamefont{Dalgarno}},
  \bibinfo{journal}{J. Phys. B: At. Mol. Opt. Phys.}
  \textbf{\bibinfo{volume}{37}}, \bibinfo{pages}{3641} (\bibinfo{year}{2004}).

\bibitem[{\citenamefont{Bodo and Gianturco}(2004)}]{bodobodo}
\bibinfo{author}{\bibfnamefont{E.}~\bibnamefont{Bodo}} \bibnamefont{and}
  \bibinfo{author}{\bibfnamefont{F.~A.} \bibnamefont{Gianturco}},
  \bibinfo{journal}{Eur. Phys. J. D.} \textbf{\bibinfo{volume}{31}},
  \bibinfo{pages}{423} (\bibinfo{year}{2004}).

\bibitem[{\citenamefont{Weck and Balakrisnan}(2004)}]{Weckeur}
\bibinfo{author}{\bibfnamefont{P.~F.} \bibnamefont{Weck}} \bibnamefont{and}
  \bibinfo{author}{\bibfnamefont{N.}~\bibnamefont{Balakrisnan}},
  \bibinfo{journal}{Eur. Phys. J. D.} \textbf{\bibinfo{volume}{31}},
  \bibinfo{pages}{417} (\bibinfo{year}{2004}).

\bibitem[{\citenamefont{Weck and Balakrisnan}(2006{\natexlab{a}})}]{WeckphyB}
\bibinfo{author}{\bibfnamefont{P.~F.} \bibnamefont{Weck}} \bibnamefont{and}
  \bibinfo{author}{\bibfnamefont{N.}~\bibnamefont{Balakrisnan}},
  \bibinfo{journal}{J. Phys. B.} \textbf{\bibinfo{volume}{39}},
  \bibinfo{pages}{S1215} (\bibinfo{year}{2006}{\natexlab{a}}).

\bibitem[{\citenamefont{Tscherbul and Krems}(2008)}]{Lihfelec}
\bibinfo{author}{\bibfnamefont{T.~V.} \bibnamefont{Tscherbul}}
  \bibnamefont{and} \bibinfo{author}{\bibfnamefont{R.~V.} \bibnamefont{Krems}},
  \bibinfo{journal}{J. Chem. Phys.} \textbf{\bibinfo{volume}{129}},
  \bibinfo{pages}{034112} (\bibinfo{year}{2008}).

\bibitem[{\citenamefont{Shapiro and Brumer}(2003)}]{Shapi}
\bibinfo{editor}{\bibfnamefont{M.}~\bibnamefont{Shapiro}} \bibnamefont{and}
  \bibinfo{editor}{\bibfnamefont{P.}~\bibnamefont{Brumer}}, eds.,
  \emph{\bibinfo{title}{Principles of the quantum control of molecular
  processes}} (\bibinfo{publisher}{John Wiley and Sons, Inc.},
  \bibinfo{year}{2003}).

\bibitem[{\citenamefont{Zeman et~al.}(2004)\citenamefont{Zeman, Shapiro, and
  Brumer}}]{Zeman}
\bibinfo{author}{\bibfnamefont{V.}~\bibnamefont{Zeman}},
  \bibinfo{author}{\bibfnamefont{M.}~\bibnamefont{Shapiro}}, \bibnamefont{and}
  \bibinfo{author}{\bibfnamefont{P.}~\bibnamefont{Brumer}},
  \bibinfo{journal}{Phys. Rev. Lett.} \textbf{\bibinfo{volume}{92}},
  \bibinfo{pages}{133204} (\bibinfo{year}{2004}).

\bibitem[{\citenamefont{Herrera}(2008)}]{FelipeH}
\bibinfo{author}{\bibfnamefont{F.}~\bibnamefont{Herrera}},
  \bibinfo{journal}{Phys. Rev. A} \textbf{\bibinfo{volume}{78}},
  \bibinfo{pages}{054702} (\bibinfo{year}{2008}).

\bibitem[{\citenamefont{Staanum et~al.}(2006)\citenamefont{Staanum, Kraft,
  Lange, Wester, and {Weidem\"{u}ller}}}]{Staanum}
\bibinfo{author}{\bibfnamefont{P.}~\bibnamefont{Staanum}},
  \bibinfo{author}{\bibfnamefont{S.~D.} \bibnamefont{Kraft}},
  \bibinfo{author}{\bibfnamefont{J.}~\bibnamefont{Lange}},
  \bibinfo{author}{\bibfnamefont{R.}~\bibnamefont{Wester}}, \bibnamefont{and}
  \bibinfo{author}{\bibfnamefont{M.}~\bibnamefont{{Weidem\"{u}ller}}},
  \bibinfo{journal}{Phys. Rev. Lett.} \textbf{\bibinfo{volume}{96}},
  \bibinfo{pages}{023201} (\bibinfo{year}{2006}).

\bibitem[{\citenamefont{Zahzam et~al.}(2006)\citenamefont{Zahzam, Vogt,
  Mudrich, Comparat, and Pillet}}]{Zahzam}
\bibinfo{author}{\bibfnamefont{N.}~\bibnamefont{Zahzam}},
  \bibinfo{author}{\bibfnamefont{T.}~\bibnamefont{Vogt}},
  \bibinfo{author}{\bibfnamefont{M.}~\bibnamefont{Mudrich}},
  \bibinfo{author}{\bibfnamefont{D.}~\bibnamefont{Comparat}}, \bibnamefont{and}
  \bibinfo{author}{\bibfnamefont{P.}~\bibnamefont{Pillet}},
  \bibinfo{journal}{Phys. Rev. Lett.} \textbf{\bibinfo{volume}{96}},
  \bibinfo{pages}{023202} (\bibinfo{year}{2006}).

\bibitem[{\citenamefont{Wynar et~al.}(2000)\citenamefont{Wynar, Freeland, Han,
  Ryu, and Heinzen}}]{Wynar}
\bibinfo{author}{\bibfnamefont{R.}~\bibnamefont{Wynar}},
  \bibinfo{author}{\bibfnamefont{R.~S.} \bibnamefont{Freeland}},
  \bibinfo{author}{\bibfnamefont{D.~J.} \bibnamefont{Han}},
  \bibinfo{author}{\bibfnamefont{C.}~\bibnamefont{Ryu}}, \bibnamefont{and}
  \bibinfo{author}{\bibfnamefont{D.~J.} \bibnamefont{Heinzen}},
  \bibinfo{journal}{Science} \textbf{\bibinfo{volume}{287}},
  \bibinfo{pages}{1016} (\bibinfo{year}{2000}).

\bibitem[{\citenamefont{Mukaiyama et~al.}(2004)\citenamefont{Mukaiyama,
  Abo-Shaeer, Xu, Chin, and Ketterle}}]{Mukaiyama}
\bibinfo{author}{\bibfnamefont{T.}~\bibnamefont{Mukaiyama}},
  \bibinfo{author}{\bibfnamefont{J.~R.} \bibnamefont{Abo-Shaeer}},
  \bibinfo{author}{\bibfnamefont{K.}~\bibnamefont{Xu}},
  \bibinfo{author}{\bibfnamefont{J.~K.} \bibnamefont{Chin}}, \bibnamefont{and}
  \bibinfo{author}{\bibfnamefont{W.}~\bibnamefont{Ketterle}},
  \bibinfo{journal}{Phys. Rev. Lett.} \textbf{\bibinfo{volume}{92}},
  \bibinfo{pages}{180402} (\bibinfo{year}{2004}).

\bibitem[{\citenamefont{Syassen et~al.}(2006)\citenamefont{Syassen, Volz,
  Teichmann, {D\"{u}rr}, and Rempe}}]{Syassen}
\bibinfo{author}{\bibfnamefont{N.}~\bibnamefont{Syassen}},
  \bibinfo{author}{\bibfnamefont{T.}~\bibnamefont{Volz}},
  \bibinfo{author}{\bibfnamefont{S.}~\bibnamefont{Teichmann}},
  \bibinfo{author}{\bibfnamefont{S.}~\bibnamefont{{D\"{u}rr}}},
  \bibnamefont{and} \bibinfo{author}{\bibfnamefont{G.}~\bibnamefont{Rempe}},
  \bibinfo{journal}{Phys. Rev. A} \textbf{\bibinfo{volume}{74}},
  \bibinfo{pages}{062706} (\bibinfo{year}{2006}).

\bibitem[{\citenamefont{Hudson et~al.}(2008)\citenamefont{Hudson, Gilfoy,
  Kotochigova, Sage, and DeMille}}]{hudsonexp}
\bibinfo{author}{\bibfnamefont{E.~R.} \bibnamefont{Hudson}},
  \bibinfo{author}{\bibfnamefont{N.~B.} \bibnamefont{Gilfoy}},
  \bibinfo{author}{\bibfnamefont{S.}~\bibnamefont{Kotochigova}},
  \bibinfo{author}{\bibfnamefont{J.~M.} \bibnamefont{Sage}}, \bibnamefont{and}
  \bibinfo{author}{\bibfnamefont{D.}~\bibnamefont{DeMille}},
  \bibinfo{journal}{Phys. Rev. Lett.} \textbf{\bibinfo{volume}{100}},
  \bibinfo{pages}{203201} (\bibinfo{year}{2008}).

\bibitem[{\citenamefont{Ospelkaus et~al.}(2010)\citenamefont{Ospelkaus, Ni,
  Wang, de~Miranda, Neyenhuis, Qu{\'{e}}m{\'{e}}ner, Julienne, Bohn, Jin, and
  Ye}}]{Ospel}
\bibinfo{author}{\bibfnamefont{S.}~\bibnamefont{Ospelkaus}},
  \bibinfo{author}{\bibfnamefont{K.-K.} \bibnamefont{Ni}},
  \bibinfo{author}{\bibfnamefont{D.}~\bibnamefont{Wang}},
  \bibinfo{author}{\bibfnamefont{M.~H.~G.} \bibnamefont{de~Miranda}},
  \bibinfo{author}{\bibfnamefont{B.}~\bibnamefont{Neyenhuis}},
  \bibinfo{author}{\bibfnamefont{G.}~\bibnamefont{Qu{\'{e}}m{\'{e}}ner}},
  \bibinfo{author}{\bibfnamefont{P.~S.} \bibnamefont{Julienne}},
  \bibinfo{author}{\bibfnamefont{J.~L.} \bibnamefont{Bohn}},
  \bibinfo{author}{\bibfnamefont{D.~S.} \bibnamefont{Jin}}, \bibnamefont{and}
  \bibinfo{author}{\bibfnamefont{J.}~\bibnamefont{Ye}},
  \bibinfo{journal}{Science} \textbf{\bibinfo{volume}{327}},
  \bibinfo{pages}{853} (\bibinfo{year}{2010}).

\bibitem[{\citenamefont{van~de Meerakker and Meijer}(2009)}]{FaraBas}
\bibinfo{author}{\bibfnamefont{S.}~\bibnamefont{van~de Meerakker}}
  \bibnamefont{and} \bibinfo{author}{\bibfnamefont{G.}~\bibnamefont{Meijer}},
  \bibinfo{journal}{Faraday Discuss.} \textbf{\bibinfo{volume}{142}},
  \bibinfo{pages}{113} (\bibinfo{year}{2009}).

\bibitem[{\citenamefont{Canosa et~al.}(2008)\citenamefont{Canosa, Goulay, Sims,
  and Rowe}}]{Canosa}
\bibinfo{author}{\bibfnamefont{A.}~\bibnamefont{Canosa}},
  \bibinfo{author}{\bibfnamefont{F.}~\bibnamefont{Goulay}},
  \bibinfo{author}{\bibfnamefont{I.~R.} \bibnamefont{Sims}}, \bibnamefont{and}
  \bibinfo{author}{\bibfnamefont{B.~R.} \bibnamefont{Rowe}},
  \emph{\bibinfo{title}{Low Temperatures and Cold Molecules}}
  (\bibinfo{publisher}{World Scientific, Singapore}, \bibinfo{year}{2008}).

\bibitem[{\citenamefont{Berteloite et~al.}(2009)\citenamefont{Berteloite, Lara,
  Picard, Dayou, Launay, Canosa, and Sims}}]{nuestroFara}
\bibinfo{author}{\bibfnamefont{C.}~\bibnamefont{Berteloite}},
  \bibinfo{author}{\bibfnamefont{M.}~\bibnamefont{Lara}},
  \bibinfo{author}{\bibfnamefont{S.~D.~L.} \bibnamefont{Picard}},
  \bibinfo{author}{\bibfnamefont{F.}~\bibnamefont{Dayou}},
  \bibinfo{author}{\bibfnamefont{J.-M.} \bibnamefont{Launay}},
  \bibinfo{author}{\bibfnamefont{A.}~\bibnamefont{Canosa}}, \bibnamefont{and}
  \bibinfo{author}{\bibfnamefont{I.~R.} \bibnamefont{Sims}},
  \bibinfo{journal}{Faraday Discuss.} \textbf{\bibinfo{volume}{142}},
  \bibinfo{pages}{236, General dicussion} (\bibinfo{year}{2009}).

\bibitem[{\citenamefont{Geppert et~al.}(2004)\citenamefont{Geppert, Goulay,
  Naulin, Costes, Canosa, Picard, and Rowe}}]{Costes}
\bibinfo{author}{\bibfnamefont{W.~D.} \bibnamefont{Geppert}},
  \bibinfo{author}{\bibfnamefont{F.}~\bibnamefont{Goulay}},
  \bibinfo{author}{\bibfnamefont{C.}~\bibnamefont{Naulin}},
  \bibinfo{author}{\bibfnamefont{M.}~\bibnamefont{Costes}},
  \bibinfo{author}{\bibfnamefont{A.}~\bibnamefont{Canosa}},
  \bibinfo{author}{\bibfnamefont{S.~D.~L.} \bibnamefont{Picard}},
  \bibnamefont{and} \bibinfo{author}{\bibfnamefont{B.~R.} \bibnamefont{Rowe}},
  \bibinfo{journal}{Phys. Chem. Chem. Phys} \textbf{\bibinfo{volume}{6}},
  \bibinfo{pages}{566} (\bibinfo{year}{2004}).

\bibitem[{\citenamefont{Costes and Naulin}(2010)}]{Costes2}
\bibinfo{author}{\bibfnamefont{M.}~\bibnamefont{Costes}} \bibnamefont{and}
  \bibinfo{author}{\bibfnamefont{C.}~\bibnamefont{Naulin}},
  \bibinfo{journal}{Phys. Chem. Chem. Phys} \textbf{\bibinfo{volume}{12}},
  \bibinfo{pages}{9154} (\bibinfo{year}{2010}).

\bibitem[{\citenamefont{Sims and Costes}(2010)}]{personal}
\bibinfo{author}{\bibfnamefont{I.~R.} \bibnamefont{Sims}} \bibnamefont{and}
  \bibinfo{author}{\bibfnamefont{M.}~\bibnamefont{Costes}}
  (\bibinfo{year}{2010}), \bibinfo{note}{personal communication}.

\bibitem[{\citenamefont{Berteloite et~al.}(2010)\citenamefont{Berteloite, Lara,
  Bergeat, Picard, Dayou, Hickson, Canosa, Naulin, Launay, Sims
  et~al.}}]{Science}
\bibinfo{author}{\bibfnamefont{C.}~\bibnamefont{Berteloite}},
  \bibinfo{author}{\bibfnamefont{M.}~\bibnamefont{Lara}},
  \bibinfo{author}{\bibfnamefont{A.}~\bibnamefont{Bergeat}},
  \bibinfo{author}{\bibfnamefont{S.~D.~L.} \bibnamefont{Picard}},
  \bibinfo{author}{\bibfnamefont{F.}~\bibnamefont{Dayou}},
  \bibinfo{author}{\bibfnamefont{K.~M.} \bibnamefont{Hickson}},
  \bibinfo{author}{\bibfnamefont{A.}~\bibnamefont{Canosa}},
  \bibinfo{author}{\bibfnamefont{C.}~\bibnamefont{Naulin}},
  \bibinfo{author}{\bibfnamefont{J.-M.} \bibnamefont{Launay}},
  \bibinfo{author}{\bibfnamefont{I.~R.} \bibnamefont{Sims}},
  \bibnamefont{et~al.} (\bibinfo{year}{2010}), \bibinfo{note}{(accepted for
  publication in Phys. Rev. Lett.)}.

\bibitem[{\citenamefont{Zyubin et~al.}(2001)\citenamefont{Zyubin, Mebel, Chao,
  and Skodje}}]{Skodje01a}
\bibinfo{author}{\bibfnamefont{A.~S.} \bibnamefont{Zyubin}},
  \bibinfo{author}{\bibfnamefont{A.~M.} \bibnamefont{Mebel}},
  \bibinfo{author}{\bibfnamefont{S.~D.} \bibnamefont{Chao}}, \bibnamefont{and}
  \bibinfo{author}{\bibfnamefont{R.~T.} \bibnamefont{Skodje}},
  \bibinfo{journal}{J. Chem. Phys.} \textbf{\bibinfo{volume}{114}},
  \bibinfo{pages}{320} (\bibinfo{year}{2001}).

\bibitem[{\citenamefont{Ho et~al.}(2002)\citenamefont{Ho, Hollebeek, Rabitz,
  Chao, Skodje, Zyubin, and Mebel}}]{Skodje02}
\bibinfo{author}{\bibfnamefont{T.-S.} \bibnamefont{Ho}},
  \bibinfo{author}{\bibfnamefont{T.}~\bibnamefont{Hollebeek}},
  \bibinfo{author}{\bibfnamefont{H.}~\bibnamefont{Rabitz}},
  \bibinfo{author}{\bibfnamefont{S.~D.} \bibnamefont{Chao}},
  \bibinfo{author}{\bibfnamefont{R.~T.} \bibnamefont{Skodje}},
  \bibinfo{author}{\bibfnamefont{A.~S.} \bibnamefont{Zyubin}},
  \bibnamefont{and} \bibinfo{author}{\bibfnamefont{A.~M.} \bibnamefont{Mebel}},
  \bibinfo{journal}{J. Chem. Phys.} \textbf{\bibinfo{volume}{116}},
  \bibinfo{pages}{4124} (\bibinfo{year}{2002}).

\bibitem[{\citenamefont{{Ba\~nares} et~al.}(2004)\citenamefont{{Ba\~nares},
  Aoiz, Honvault, and Launay}}]{Banares02}
\bibinfo{author}{\bibfnamefont{L.}~\bibnamefont{{Ba\~nares}}},
  \bibinfo{author}{\bibfnamefont{F.}~\bibnamefont{Aoiz}},
  \bibinfo{author}{\bibfnamefont{P.}~\bibnamefont{Honvault}}, \bibnamefont{and}
  \bibinfo{author}{\bibfnamefont{J.-M.} \bibnamefont{Launay}},
  \bibinfo{journal}{J. Phys. Chem. A} \textbf{\bibinfo{volume}{108}},
  \bibinfo{pages}{1616} (\bibinfo{year}{2004}).

\bibitem[{\citenamefont{{Ba\~nares} et~al.}(2005)\citenamefont{{Ba\~nares},
  Castillo, Honvault, and Launay}}]{Banaresulti}
\bibinfo{author}{\bibfnamefont{L.}~\bibnamefont{{Ba\~nares}}},
  \bibinfo{author}{\bibfnamefont{J.~F.} \bibnamefont{Castillo}},
  \bibinfo{author}{\bibfnamefont{P.}~\bibnamefont{Honvault}}, \bibnamefont{and}
  \bibinfo{author}{\bibfnamefont{J.-M.} \bibnamefont{Launay}},
  \bibinfo{journal}{Phys. Chem. Chem. Phys.} \textbf{\bibinfo{volume}{7}},
  \bibinfo{pages}{627} (\bibinfo{year}{2005}).

\bibitem[{\citenamefont{Honvault and Launay}(2003)}]{HonvaLau}
\bibinfo{author}{\bibfnamefont{P.}~\bibnamefont{Honvault}} \bibnamefont{and}
  \bibinfo{author}{\bibfnamefont{J.-M.} \bibnamefont{Launay}},
  \bibinfo{journal}{Chem. Phys. Lett.} \textbf{\bibinfo{volume}{370}},
  \bibinfo{pages}{371} (\bibinfo{year}{2003}).

\bibitem[{\citenamefont{Rackham et~al.}(2003)\citenamefont{Rackham,
  Gonz\'{a}lez-Lezana, and Manolopoulos}}]{Rack1}
\bibinfo{author}{\bibfnamefont{E.~J.} \bibnamefont{Rackham}},
  \bibinfo{author}{\bibfnamefont{T.}~\bibnamefont{Gonz\'{a}lez-Lezana}},
  \bibnamefont{and} \bibinfo{author}{\bibfnamefont{D.~E.}
  \bibnamefont{Manolopoulos}}, \bibinfo{journal}{J. Chem. Phys.}
  \textbf{\bibinfo{volume}{119}}, \bibinfo{pages}{12895}
  (\bibinfo{year}{2003}).

\bibitem[{\citenamefont{Gonz\'{a}lez-Lezana}(2007)}]{Tomas}
\bibinfo{author}{\bibfnamefont{T.}~\bibnamefont{Gonz\'{a}lez-Lezana}},
  \bibinfo{journal}{Int. Rev. Phys. Chem.} \textbf{\bibinfo{volume}{26}},
  \bibinfo{pages}{29} (\bibinfo{year}{2007}).

\bibitem[{\citenamefont{Klos et~al.}(2007)\citenamefont{Klos, Dagdigian, and
  Alexander}}]{klos}
\bibinfo{author}{\bibfnamefont{J.~A.} \bibnamefont{Klos}},
  \bibinfo{author}{\bibfnamefont{P.~J.} \bibnamefont{Dagdigian}},
  \bibnamefont{and} \bibinfo{author}{\bibfnamefont{M.~H.}
  \bibnamefont{Alexander}}, \bibinfo{journal}{J. Chem. Phys.}
  \textbf{\bibinfo{volume}{127}}, \bibinfo{pages}{154321}
  (\bibinfo{year}{2007}).

\bibitem[{\citenamefont{Maiti et~al.}(2004)\citenamefont{Maiti, Schatz, and
  Lendvay}}]{Maiti}
\bibinfo{author}{\bibfnamefont{B.}~\bibnamefont{Maiti}},
  \bibinfo{author}{\bibfnamefont{G.~C.} \bibnamefont{Schatz}},
  \bibnamefont{and} \bibinfo{author}{\bibfnamefont{G.}~\bibnamefont{Lendvay}},
  \bibinfo{journal}{J. Phys. Chem. A} \textbf{\bibinfo{volume}{108}},
  \bibinfo{pages}{8772} (\bibinfo{year}{2004}).

\bibitem[{\citenamefont{Lee and Liu}(1998{\natexlab{a}})}]{Liu98a}
\bibinfo{author}{\bibfnamefont{S.-H.} \bibnamefont{Lee}} \bibnamefont{and}
  \bibinfo{author}{\bibfnamefont{K.}~\bibnamefont{Liu}}, \bibinfo{journal}{J.
  Phys. Chem.} \textbf{\bibinfo{volume}{102}}, \bibinfo{pages}{8637}
  (\bibinfo{year}{1998}{\natexlab{a}}).

\bibitem[{\citenamefont{Lee and Liu}(1998{\natexlab{b}})}]{Liu98b}
\bibinfo{author}{\bibfnamefont{S.-H.} \bibnamefont{Lee}} \bibnamefont{and}
  \bibinfo{author}{\bibfnamefont{K.}~\bibnamefont{Liu}},
  \bibinfo{journal}{Chem. Phys. Lett.} \textbf{\bibinfo{volume}{290}},
  \bibinfo{pages}{323} (\bibinfo{year}{1998}{\natexlab{b}}).

\bibitem[{\citenamefont{Lee and Liu}(2000)}]{Liu00}
\bibinfo{author}{\bibfnamefont{S.-H.} \bibnamefont{Lee}} \bibnamefont{and}
  \bibinfo{author}{\bibfnamefont{K.}~\bibnamefont{Liu}},
  \emph{\bibinfo{title}{in Advances in Molecular Beam Research and
  Applications}} (\bibinfo{publisher}{Springer-Verlag, Berlin},
  \bibinfo{year}{2000}).

\bibitem[{\citenamefont{Aoiz et~al.}(2006)\citenamefont{Aoiz, {Ba\~nares}, and
  Herrero}}]{aoizfe}
\bibinfo{author}{\bibfnamefont{F.}~\bibnamefont{Aoiz}},
  \bibinfo{author}{\bibfnamefont{L.}~\bibnamefont{{Ba\~nares}}},
  \bibnamefont{and} \bibinfo{author}{\bibfnamefont{V.~J.}
  \bibnamefont{Herrero}}, \bibinfo{journal}{J. Phys. Chem. A}
  \textbf{\bibinfo{volume}{110}}, \bibinfo{pages}{12546}
  (\bibinfo{year}{2006}).

\bibitem[{\citenamefont{A.~H. H.~Chang}(2000)}]{Changg}
\bibinfo{author}{\bibfnamefont{S.~H.~L.} \bibnamefont{A.~H. H.~Chang}},
  \bibinfo{journal}{Chem. Phys. Lett.} \textbf{\bibinfo{volume}{320}},
  \bibinfo{pages}{161} (\bibinfo{year}{2000}).

\bibitem[{\citenamefont{Aoiz et~al.}(2008)\citenamefont{Aoiz,
  Gonz\'{a}lez-Lezana, and R\'{a}banos}}]{Aoiztomas}
\bibinfo{author}{\bibfnamefont{F.}~\bibnamefont{Aoiz}},
  \bibinfo{author}{\bibfnamefont{T.}~\bibnamefont{Gonz\'{a}lez-Lezana}},
  \bibnamefont{and} \bibinfo{author}{\bibfnamefont{V.~S.}
  \bibnamefont{R\'{a}banos}}, \bibinfo{journal}{J. Chem. Phys.}
  \textbf{\bibinfo{volume}{129}}, \bibinfo{pages}{094305}
  (\bibinfo{year}{2008}).

\bibitem[{\citenamefont{Lin and Guo}(2005)}]{Ying}
\bibinfo{author}{\bibfnamefont{S.~Y.} \bibnamefont{Lin}} \bibnamefont{and}
  \bibinfo{author}{\bibfnamefont{H.}~\bibnamefont{Guo}}, \bibinfo{journal}{J.
  Chem. Phys.} \textbf{\bibinfo{volume}{122}}, \bibinfo{pages}{074304}
  (\bibinfo{year}{2005}).

\bibitem[{\citenamefont{Rackham et~al.}(2001)\citenamefont{Rackham,
  Huarte-Larranaga, and Manolopoulos}}]{Rack0}
\bibinfo{author}{\bibfnamefont{E.~J.} \bibnamefont{Rackham}},
  \bibinfo{author}{\bibfnamefont{F.}~\bibnamefont{Huarte-Larranaga}},
  \bibnamefont{and} \bibinfo{author}{\bibfnamefont{D.~E.}
  \bibnamefont{Manolopoulos}}, \bibinfo{journal}{Chem. Phys. Lett.}
  \textbf{\bibinfo{volume}{343}}, \bibinfo{pages}{356} (\bibinfo{year}{2001}).

\bibitem[{\citenamefont{Aoiz et~al.}(2007)\citenamefont{Aoiz, R\'{a}banos,
  Gonz\'{a}lez-Lezana, and Manolopoulos}}]{Aoiz3}
\bibinfo{author}{\bibfnamefont{F.}~\bibnamefont{Aoiz}},
  \bibinfo{author}{\bibfnamefont{V.~S.} \bibnamefont{R\'{a}banos}},
  \bibinfo{author}{\bibfnamefont{T.}~\bibnamefont{Gonz\'{a}lez-Lezana}},
  \bibnamefont{and} \bibinfo{author}{\bibfnamefont{D.~E.}
  \bibnamefont{Manolopoulos}}, \bibinfo{journal}{J. Chem. Phys.}
  \textbf{\bibinfo{volume}{126}}, \bibinfo{pages}{161101}
  (\bibinfo{year}{2007}).

\bibitem[{\citenamefont{Chu et~al.}(2007)\citenamefont{Chu, Han, and
  Schatz}}]{Tian}
\bibinfo{author}{\bibfnamefont{T.-S.} \bibnamefont{Chu}},
  \bibinfo{author}{\bibfnamefont{K.-L.} \bibnamefont{Han}}, \bibnamefont{and}
  \bibinfo{author}{\bibfnamefont{G.~C.} \bibnamefont{Schatz}},
  \bibinfo{journal}{J. Phys. Chem. A} \textbf{\bibinfo{volume}{111}},
  \bibinfo{pages}{8286} (\bibinfo{year}{2007}).

\bibitem[{\citenamefont{Weck and Balakrisnan}(2006{\natexlab{b}})}]{Weck2006}
\bibinfo{author}{\bibfnamefont{P.~F.} \bibnamefont{Weck}} \bibnamefont{and}
  \bibinfo{author}{\bibfnamefont{N.}~\bibnamefont{Balakrisnan}},
  \bibinfo{journal}{Int. Rev. Phys. Chem.} \textbf{\bibinfo{volume}{25}},
  \bibinfo{pages}{283} (\bibinfo{year}{2006}{\natexlab{b}}).

\bibitem[{\citenamefont{Launay and Dourneuf}(1990)}]{Launayfirst}
\bibinfo{author}{\bibfnamefont{J.~M.} \bibnamefont{Launay}} \bibnamefont{and}
  \bibinfo{author}{\bibfnamefont{M.~L.} \bibnamefont{Dourneuf}},
  \bibinfo{journal}{Chem. Phys. Lett} \textbf{\bibinfo{volume}{169}},
  \bibinfo{pages}{473} (\bibinfo{year}{1990}).

\bibitem[{\citenamefont{Bussery-Honvault
  et~al.}(2008)\citenamefont{Bussery-Honvault, Dayou, and
  Zanchet}}]{bussery:08}
\bibinfo{author}{\bibfnamefont{B.}~\bibnamefont{Bussery-Honvault}},
  \bibinfo{author}{\bibfnamefont{F.}~\bibnamefont{Dayou}}, \bibnamefont{and}
  \bibinfo{author}{\bibfnamefont{A.}~\bibnamefont{Zanchet}},
  \bibinfo{journal}{J. Chem. Phys.} \textbf{\bibinfo{volume}{129}},
  \bibinfo{pages}{234302} (\bibinfo{year}{2008}).

\bibitem[{\citenamefont{Groenenboom et~al.}(2007)\citenamefont{Groenenboom,
  Chu, and Krems}}]{groenenboom:06}
\bibinfo{author}{\bibfnamefont{G.~C.} \bibnamefont{Groenenboom}},
  \bibinfo{author}{\bibfnamefont{X.}~\bibnamefont{Chu}}, \bibnamefont{and}
  \bibinfo{author}{\bibfnamefont{R.~V.} \bibnamefont{Krems}},
  \bibinfo{journal}{J. Chem. Phys.} \textbf{\bibinfo{volume}{126}},
  \bibinfo{pages}{204306} (\bibinfo{year}{2007}).

\bibitem[{\citenamefont{Spelsberg}(1999)}]{spelsberg:99}
\bibinfo{author}{\bibfnamefont{D.}~\bibnamefont{Spelsberg}},
  \bibinfo{journal}{J. Chem. Phys.} \textbf{\bibinfo{volume}{111}},
  \bibinfo{pages}{9625} (\bibinfo{year}{1999}).

\bibitem[{\citenamefont{Zeimen et~al.}(2003)\citenamefont{Zeimen, os,
  Groenenboom, and van~der Avoird}}]{zeimen:03}
\bibinfo{author}{\bibfnamefont{W.~B.} \bibnamefont{Zeimen}},
  \bibinfo{author}{\bibfnamefont{J.~K.} \bibnamefont{os}},
  \bibinfo{author}{\bibfnamefont{G.~C.} \bibnamefont{Groenenboom}},
  \bibnamefont{and} \bibinfo{author}{\bibfnamefont{A.}~\bibnamefont{van~der
  Avoird}}, \bibinfo{journal}{J. chem. Phys.} \textbf{\bibinfo{volume}{118}},
  \bibinfo{pages}{7340} (\bibinfo{year}{2003}).

\bibitem[{\citenamefont{Spelsberg et~al.}(1993)\citenamefont{Spelsberg, Lorenz,
  and Meyer}}]{spelsberg:93}
\bibinfo{author}{\bibfnamefont{D.}~\bibnamefont{Spelsberg}},
  \bibinfo{author}{\bibfnamefont{T.}~\bibnamefont{Lorenz}}, \bibnamefont{and}
  \bibinfo{author}{\bibfnamefont{W.}~\bibnamefont{Meyer}}, \bibinfo{journal}{J.
  Chem. Phys.} \textbf{\bibinfo{volume}{99}}, \bibinfo{pages}{7845}
  (\bibinfo{year}{1993}).

\bibitem[{\citenamefont{Bishop and Pipin}(1992)}]{bishop:92}
\bibinfo{author}{\bibfnamefont{D.~M.} \bibnamefont{Bishop}} \bibnamefont{and}
  \bibinfo{author}{\bibfnamefont{J.}~\bibnamefont{Pipin}}, \bibinfo{journal}{J.
  Chem. Phys.} \textbf{\bibinfo{volume}{97}}, \bibinfo{pages}{3375}
  (\bibinfo{year}{1992}).

\bibitem[{\citenamefont{Electronic~structure program}(2005)}]{dalton}
\bibinfo{author}{\bibfnamefont{R.~.~h.} \bibnamefont{Electronic~structure
  program}} (\bibinfo{year}{2005}).

\bibitem[{\citenamefont{Medved et~al.}(2000)\citenamefont{Medved, Fowler, and
  Hutson}}]{medved:00}
\bibinfo{author}{\bibfnamefont{M.}~\bibnamefont{Medved}},
  \bibinfo{author}{\bibfnamefont{P.~W.} \bibnamefont{Fowler}},
  \bibnamefont{and} \bibinfo{author}{\bibfnamefont{J.~M.}
  \bibnamefont{Hutson}}, \bibinfo{journal}{Mol. Phys.}
  \textbf{\bibinfo{volume}{98}}, \bibinfo{pages}{453} (\bibinfo{year}{2000}).

\bibitem[{\citenamefont{Andersson and Sadlej}(1992)}]{andersson:92}
\bibinfo{author}{\bibfnamefont{K.}~\bibnamefont{Andersson}} \bibnamefont{and}
  \bibinfo{author}{\bibfnamefont{A.~J.} \bibnamefont{Sadlej}},
  \bibinfo{journal}{Phys. Rev. A} \textbf{\bibinfo{volume}{46}},
  \bibinfo{pages}{2356} (\bibinfo{year}{1992}).

\bibitem[{\citenamefont{Langhoff and Karplus}(1970)}]{langhoff:70}
\bibinfo{author}{\bibfnamefont{P.~W.} \bibnamefont{Langhoff}} \bibnamefont{and}
  \bibinfo{author}{\bibfnamefont{M.}~\bibnamefont{Karplus}},
  \bibinfo{journal}{J. Chem. Phys.} \textbf{\bibinfo{volume}{53}},
  \bibinfo{pages}{233} (\bibinfo{year}{1970}).

\bibitem[{\citenamefont{Sold\'{a}n et~al.}(2002)\citenamefont{Sold\'{a}n,
  Cvita\v{s}, Hutson, Honvault, and Launay}}]{Sol02}
\bibinfo{author}{\bibfnamefont{P.}~\bibnamefont{Sold\'{a}n}},
  \bibinfo{author}{\bibfnamefont{M.~T.} \bibnamefont{Cvita\v{s}}},
  \bibinfo{author}{\bibfnamefont{J.~M.} \bibnamefont{Hutson}},
  \bibinfo{author}{\bibfnamefont{P.}~\bibnamefont{Honvault}}, \bibnamefont{and}
  \bibinfo{author}{\bibfnamefont{J.-M.} \bibnamefont{Launay}},
  \bibinfo{journal}{Phys. Rev. Lett.} \textbf{\bibinfo{volume}{89}},
  \bibinfo{pages}{153201} (\bibinfo{year}{2002}).

\bibitem[{\citenamefont{Qu{\'{e}}m{\'{e}}ner
  et~al.}(2004)\citenamefont{Qu{\'{e}}m{\'{e}}ner, Honvault, and
  Launay}}]{Quem04}
\bibinfo{author}{\bibfnamefont{G.}~\bibnamefont{Qu{\'{e}}m{\'{e}}ner}},
  \bibinfo{author}{\bibfnamefont{P.}~\bibnamefont{Honvault}}, \bibnamefont{and}
  \bibinfo{author}{\bibfnamefont{J.-M.} \bibnamefont{Launay}},
  \bibinfo{journal}{Eur. Phys. J. D.} \textbf{\bibinfo{volume}{30}},
  \bibinfo{pages}{201} (\bibinfo{year}{2004}).

\bibitem[{\citenamefont{Honvault and Dynamics}(2004)}]{hon04}
\bibinfo{author}{\bibfnamefont{P.}~\bibnamefont{Honvault}} \bibnamefont{and}
  \bibinfo{author}{\bibfnamefont{J.-M.~L.} \bibnamefont{Dynamics}},
  \emph{\bibinfo{title}{in {\sl Theory of Chemical Reaction Dynamics}}}
  (\bibinfo{publisher}{NATO Science Series vol. 145, Kluwer},
  \bibinfo{year}{2004}).

\bibitem[{\citenamefont{Lester}(1971)}]{Lester}
\bibinfo{author}{\bibfnamefont{W.~A.} \bibnamefont{Lester}},
  \bibinfo{journal}{Meth. Comput. Phys.} \textbf{\bibinfo{volume}{10}},
  \bibinfo{pages}{211} (\bibinfo{year}{1971}).

\bibitem[{\citenamefont{Song and Varandas}(2009)}]{Var}
\bibinfo{author}{\bibfnamefont{Y.~Z.} \bibnamefont{Song}} \bibnamefont{and}
  \bibinfo{author}{\bibfnamefont{A.~J.~C.} \bibnamefont{Varandas}},
  \bibinfo{journal}{J. Chem. Phys.} \textbf{\bibinfo{volume}{130}},
  \bibinfo{pages}{134317} (\bibinfo{year}{2009}).

\bibitem[{\citenamefont{Drukker}(1999)}]{Druk}
\bibinfo{author}{\bibfnamefont{G.~C. S.~K.} \bibnamefont{Drukker}},
  \bibinfo{journal}{J. Chem. Phys.} \textbf{\bibinfo{volume}{111}},
  \bibinfo{pages}{2451} (\bibinfo{year}{1999}).

\bibitem[{\citenamefont{Black and Jusinski}(1986)}]{Jusin}
\bibinfo{author}{\bibfnamefont{G.}~\bibnamefont{Black}} \bibnamefont{and}
  \bibinfo{author}{\bibfnamefont{L.~E.} \bibnamefont{Jusinski}},
  \bibinfo{journal}{Chem. Phys. Lett.} \textbf{\bibinfo{volume}{124}},
  \bibinfo{pages}{90} (\bibinfo{year}{1986}).

\bibitem[{\citenamefont{Levine and Bernstein}(1987)}]{Lange}
\bibinfo{author}{\bibfnamefont{R.~D.} \bibnamefont{Levine}} \bibnamefont{and}
  \bibinfo{author}{\bibfnamefont{R.~B.} \bibnamefont{Bernstein}},
  \emph{\bibinfo{title}{in {\sl Molecular Reaction Dynamics and Chemical
  Reactivity}}} (\bibinfo{publisher}{Oxford University Press},
  \bibinfo{year}{1987}).

\bibitem[{\citenamefont{Qu{\'{e}}m{\'{e}}ner
  et~al.}(2005)\citenamefont{Qu{\'{e}}m{\'{e}}ner, Honvault, Launay,
  Sold{\'{a}}n, Potter, and Hutson}}]{Quem05}
\bibinfo{author}{\bibfnamefont{G.}~\bibnamefont{Qu{\'{e}}m{\'{e}}ner}},
  \bibinfo{author}{\bibfnamefont{P.}~\bibnamefont{Honvault}},
  \bibinfo{author}{\bibfnamefont{J.-M.} \bibnamefont{Launay}},
  \bibinfo{author}{\bibfnamefont{P.}~\bibnamefont{Sold{\'{a}}n}},
  \bibinfo{author}{\bibfnamefont{D.}~\bibnamefont{Potter}}, \bibnamefont{and}
  \bibinfo{author}{\bibfnamefont{J.~M.} \bibnamefont{Hutson}},
  \bibinfo{journal}{Phys. Rev. A} \textbf{\bibinfo{volume}{71}},
  \bibinfo{pages}{032722} (\bibinfo{year}{2005}).

\bibitem[{\citenamefont{Cvita\v{s} et~al.}(2005)\citenamefont{Cvita\v{s},
  Sold\'{a}n, Hutson, Honvault, and Launay}}]{SolPRL}
\bibinfo{author}{\bibfnamefont{M.~T.} \bibnamefont{Cvita\v{s}}},
  \bibinfo{author}{\bibfnamefont{P.}~\bibnamefont{Sold\'{a}n}},
  \bibinfo{author}{\bibfnamefont{J.~M.} \bibnamefont{Hutson}},
  \bibinfo{author}{\bibfnamefont{P.}~\bibnamefont{Honvault}}, \bibnamefont{and}
  \bibinfo{author}{\bibfnamefont{J.-M.} \bibnamefont{Launay}},
  \bibinfo{journal}{Phys. Rev. Lett.} \textbf{\bibinfo{volume}{94}},
  \bibinfo{pages}{033201} (\bibinfo{year}{2005}).

\bibitem[{\citenamefont{Qu{\'{e}}m{\'{e}}ner and Bohn}(2010)}]{QueB}
\bibinfo{author}{\bibfnamefont{G.}~\bibnamefont{Qu{\'{e}}m{\'{e}}ner}}
  \bibnamefont{and} \bibinfo{author}{\bibfnamefont{J.~L.} \bibnamefont{Bohn}},
  \bibinfo{journal}{Phys. Rev. A} \textbf{\bibinfo{volume}{81}},
  \bibinfo{pages}{022702} (\bibinfo{year}{2010}).

\bibitem[{\citenamefont{Balakrishnan}(2004)}]{Balak}
\bibinfo{author}{\bibfnamefont{N.}~\bibnamefont{Balakrishnan}},
  \bibinfo{journal}{J. Chem. Phys.} \textbf{\bibinfo{volume}{121}},
  \bibinfo{pages}{5563} (\bibinfo{year}{2004}).

\bibitem[{\citenamefont{Weck and Balakrisnan}(2005)}]{WeckLihf}
\bibinfo{author}{\bibfnamefont{P.~F.} \bibnamefont{Weck}} \bibnamefont{and}
  \bibinfo{author}{\bibfnamefont{N.}~\bibnamefont{Balakrisnan}},
  \bibinfo{journal}{J. Chem. Phys.} \textbf{\bibinfo{volume}{122}},
  \bibinfo{pages}{154309} (\bibinfo{year}{2005}).

\bibitem[{\citenamefont{Bodo et~al.}(2002)\citenamefont{Bodo, Gianturco, and
  Dalgarno}}]{bodoG}
\bibinfo{author}{\bibfnamefont{E.}~\bibnamefont{Bodo}},
  \bibinfo{author}{\bibfnamefont{F.~A.} \bibnamefont{Gianturco}},
  \bibnamefont{and} \bibinfo{author}{\bibfnamefont{A.}~\bibnamefont{Dalgarno}},
  \bibinfo{journal}{J. Chem. Phys.} \textbf{\bibinfo{volume}{116}},
  \bibinfo{pages}{9222} (\bibinfo{year}{2002}).

\bibitem[{\citenamefont{Qiu et~al.}(2006)\citenamefont{Qiu, Ren, Che, Dai,
  Harich, Wang, Yang, Xu, Xie, Gustafsson et~al.}}]{Xueming}
\bibinfo{author}{\bibfnamefont{M.}~\bibnamefont{Qiu}},
  \bibinfo{author}{\bibfnamefont{Z.}~\bibnamefont{Ren}},
  \bibinfo{author}{\bibfnamefont{L.}~\bibnamefont{Che}},
  \bibinfo{author}{\bibfnamefont{D.}~\bibnamefont{Dai}},
  \bibinfo{author}{\bibfnamefont{S.~A.} \bibnamefont{Harich}},
  \bibinfo{author}{\bibfnamefont{X.}~\bibnamefont{Wang}},
  \bibinfo{author}{\bibfnamefont{X.}~\bibnamefont{Yang}},
  \bibinfo{author}{\bibfnamefont{C.}~\bibnamefont{Xu}},
  \bibinfo{author}{\bibfnamefont{D.}~\bibnamefont{Xie}},
  \bibinfo{author}{\bibfnamefont{M.}~\bibnamefont{Gustafsson}},
  \bibnamefont{et~al.}, \bibinfo{journal}{Science}
  \textbf{\bibinfo{volume}{311}}, \bibinfo{pages}{1440} (\bibinfo{year}{2006}).

\bibitem[{\citenamefont{Schatz}(2000)}]{Schatz}
\bibinfo{author}{\bibfnamefont{G.~C.} \bibnamefont{Schatz}},
  \bibinfo{journal}{Nature} \textbf{\bibinfo{volume}{288}},
  \bibinfo{pages}{1599} (\bibinfo{year}{2000}).

\bibitem[{\citenamefont{Skodje et~al.}(2000)\citenamefont{Skodje, Skouteris,
  Manolopoulos, Lee, Dong, and Liu}}]{Skodje}
\bibinfo{author}{\bibfnamefont{R.~T.} \bibnamefont{Skodje}},
  \bibinfo{author}{\bibfnamefont{D.}~\bibnamefont{Skouteris}},
  \bibinfo{author}{\bibfnamefont{D.~E.} \bibnamefont{Manolopoulos}},
  \bibinfo{author}{\bibfnamefont{S.-H.} \bibnamefont{Lee}},
  \bibinfo{author}{\bibfnamefont{F.}~\bibnamefont{Dong}}, \bibnamefont{and}
  \bibinfo{author}{\bibfnamefont{K.}~\bibnamefont{Liu}},
  \bibinfo{journal}{Phys. Rev. Lett.} \textbf{\bibinfo{volume}{85}},
  \bibinfo{pages}{1206} (\bibinfo{year}{2000}).

\end{thebibliography}

\end{document}